\def\lae{\mathrel{<\kern-1.0em\lower0.9ex\hbox{$\sim$}}}
\def\gae{\mathrel{>\kern-1.0em\lower0.9ex\hbox{$\sim$}}}

\newcommand{\be}{\begin{equation}}
\newcommand{\ee}{\end{equation}}

\documentclass[12pt,preprint]{aastex}
\usepackage{rotating}
\usepackage{amsmath}
\shorttitle{Formation of the Electronic Spectrum} \shortauthors{Zheng et al.}

\begin{document}

\title{Formation of the Electronic Spectrum in Relativistic Jets of Gamma-Ray Blazars}
\author{Y.G. Zheng\altaffilmark{1,2,3,4}, G.B. Long\altaffilmark{5}; C.Y. Yang\altaffilmark{3,4}; J.M. Bai\altaffilmark{3,4};}
\altaffiltext{1}{Department of Physics, Yunnan Normal University, Kunming 650092, China (E-mail:ynzyg@ynu.edu.cn)}
\altaffiltext{2}{Shandong Provincial Key Laboratory of Optical Astronomy and Solar-Terrestrial Environment£¬Shandong University, Weihai, 264209}
\altaffiltext{3}{Yunnan Observatories, Chinese Academy of Sciences, Kunming 650011, China (E-mail:baijinming@ynao.ac.cn)}
\altaffiltext{4}{Key Laboratory for the Structure and Evolution of Celestial Objects, Chinese Academy of Sciences }
\altaffiltext{5}{Department of Physics and Astronomy, Sun Yat-Sen University, Zhuhai 519082, China}

\begin{abstract}
We describe the theory of \emph{Fermi}-type acceleration, including first-order \emph{Fermi} acceleration at a parallel shock front and second-order \emph{Fermi} acceleration in a test particle limit. Including the theory of the turbulent acceleration and the derivation of the general Fokker-Planck equation, we take into account the basic particle-transport equation and the Fokker-Planck equation with spatially homogeneous isotropic distribution. In the cases of some special physical processes, we construct the particle-transport equations, and compile the analytical or semi-analytical solutions in 11 different cases. Even though, traditionally, one of the electron energy distributions for these different cases can be used to reproduce the multi-wavelength emission of a blazar, due to very long integration times of high-energy observations, a time-integrated electron energy distribution should be introduced in approaches to modeling the multi-wavelength spectral energy distribution of blazars, especially in modeling the flaring state.
\end{abstract}





\section{Introduction}

A blazar is a special class of active galactic nucleus (AGN) with a non-thermal continuum emission that arises from the jet emission taking place in an AGN whose jet axis is closely aligned with the observer's line of sight (Urry and Padovani 1995). In general, the multi-wavelength spectral energy distribution (SED) of a blazar is characterized by two humps, indicating two components (Fossati \textit{et al}. 1998). It is widely admitted that the low-energy hump that extends from the radio to ultraviolet (UV)/X-ray regime is produced by synchrotron radiation from relativistic electrons in the jet, although the high-energy hump that extends from the X-ray to the high-energy $\gamma$-ray regime remains an open issue. In the lepton model scenarios, the high-energy hump is attributed to inverse-Compton (IC) scattering of seed photons.

\emph{Fermi}-type acceleration, including first-order \emph{Fermi} acceleration at a shock front (so-called diffusive shock acceleration) and second-order \emph{Fermi} acceleration by plasma turbulence (so-called stochastic acceleration) has been widely applied to explain different kinds of high-energy astrophysical phenomena. Non-relativistic diffusive shock acceleration (DSA) is a relatively efficient acceleration mechanism (Vannoni \textit{et al}. 2009; Virtanen \textit{et al}. 2005) whose process can produce a power-law energy spectrum (e.g. Krymsky \textit{et al}. 1977; Axford \textit{et al}. 1977; Bell 1978; Drury 1983; Blandford \textit{et al}. 1987; Jones \textit{et al.} 1991; Protheroe 2004). However, in the relativistic case, the highly anisotropic particle distribution near the shock results in more complicated physics (e.g., Bednarz and Ostrowski 1998; Kirk \textit{et al}. 2000; Achterberg \textit{et al}. 2001; Keshet\textit{ et al}. 2005; Ellison\textit{ et al}. 2013; Spitkovsky 2008; Sironi \textit{et al}. 2015; Schlickeiser 2015). The stochastic acceleration (SA) is also an efficient acceleration mechanism which can produce a hard spectrum (e.g. Schlickeiser 1984; 1985; Park \textit{et al.} 1995; Becker \textit{et al}. (2006), Stawarz \textit{et al}. (2008). The SA has been investigated by the authors such as Skilling (1975), Schlickeiser (1989), Miller and Roberts (1995), Schlickeiser (2002), Petrosian (2004), and Dermer \textit{et al.} (2009) and has been successfully applied to blazars (e.g. Schlickeiser and Dermer 2000; Katarzynski \textit{et al}. 2006; Zheng e\textit{t al}. 2011a; Asano \textit{et al}. 2014).

The leptonic model was widely applied to model the SED of a blazar. In the framework of the standard blazar paradigm, the SED can be reproduced by either a synchrotron-self-Compton (SSC) process or external-Compton (EC) process, or both, in which a non-thermal relativistic electron energy distribution (EED) is assumed according to the observed spectrum characteristics. Various types of  EEDs are adopted in the main literature,  including a power law (e.g., Katarzynski \textit{et al.} 2006), broken power law (e.g., Tavecchio \textit{et al}. 1998; Katarzynski \textit{et al.} 2001; Albert \textit{et al}. 2007; Tavecchio \textit{et al}. 2010), log parabolic (e.g., Massaro \textit{et al}. 2004; Tramacere \textit{et al}. 2011; Hayashida \textit{et al}. 2012; Chen \textit{et al.} 2014 ), power law with an exponential high-energy cutoff (e.g., Finke \textit{et al.} 2008; Yan \textit{et al}. 2013), power law at low energies with a log-parabolic high-energy branch (e.g., Massaro \textit{et al}. 2006; Tramacere \textit{et al}. 2009), double broken power law (e.g., Abdo \textit{et al}. 2011a; 2011b), and etc. The non-thermal EEDs that are required by the observed SED properties of a jet results from \emph{Fermi} processes (Rieger \textit{et al}. 2007; Paggi 2010; Tramacere \textit{et al.} 2011; Abdo \textit{et al}. 2011a; 2011b).

The shape of an EED is determined by the particle kinetic equation, which includes \emph{Fermi} acceleration and other physical processes, such as injection, radiative cooling, and escapes. However, the EED encodes important information about the jet physics. In principle, constructing the particle-transport equation and obtaining its solution can reproduce observed spectra in the standard blazar paradigm, and then invert the physical information of the source. In these scenarios, the EED in the standard blazar paradigm can invert the physical process in the blazar region. Since the analytical solution of the particle-transport equation more easily provides insight into the inner physical mechanism than the numerical solution, many authors (e.g., Kardashev 1962; Schlickeiser 1984; 1985; Park \textit{et al}. 1995;  Kirk \textit{et al}. 1998; Keshet \textit{et al}. 2005; Becker \textit{et al}. 2006; Stawarz \textit{et al}. 2008; Dermer \textit{et al}. 2009; Tramacere \textit{et al}. 2009; Mertsch \textit{et al}. 2011; Finke \textit{et al}. 2013) try to solve the second-order Fokker-Planck or the first-order particle-transport equation in certain conditions in an analytical or semi-analytical way.

In order to invert the jet physics, in this paper we reconstruct the particle-transport equation in a general way and briefly derive  all the analytical solutions and separately explain their physical implications. The aim of the present work is to study in more detail the solution of the energy or momentum diffusion equations in the different physical conditions, in which the stochastic \emph{Fermi} acceleration, radiative losses, particle injection, and escapes are contained.

\section{Introduction to basic physics of particle acceleration}
{\bf\emph{Fermi Acceleration}}~~The particle acceleration process is that in which the non-thermal particles distract energy from the kinetic energy of bulk flow. Both the stochastic and shock \emph{Fermi} mechanisms belong to statistical acceleration. These mechanisms rely on scattering test particles by moving scattering centers; that is, magnetic inhomogeneities or plasma turbulence waves that are probably excited via Kelvin-Helmholtz instability and/or the current-driven instability. Such instabilities may be triggered by jet recollimations that are induced by a pressure gradient in the medium (Asano \textit{et al}. 2014). In the first-order \emph{Fermi} process, the particles' motions are described statistically crossing the shock discontinuity back and forth many times by being elastically scattered between approaching irregularities or plasma waves upstream and downstream of the shock particles. The particles gain energy that is proportional to the shock velocity per shock crossing. The second-order \emph{Fermi} acceleration assumes elastic scattering of the charged particles by moving scattering centers, leading to a diffusive energy gain. Because energy-gaining head-on collisions are more numerous than energy-losing trailing ones, the net energy gain per bounce (or the acceleration rate that is usually relative to the spectrum index of accelerated particles) is proportional to the square of the scattering center velocity (Lee 1982; Malkov \textit{et al}. 2001; Katarzynski \textit{et al}. 2006). To reach the maximum energy and be able to confine it within the accelerator, the particles that are accelerated should be able to efficiently resonate with the scattering centers in the \emph{Fermi} mechanism (Protheroe \textit{et al}. 2004). Actually, the time in DSA for the accelerated particles to achieve one cycle should be longer than the the time in SA, and the acceleration rate of DSA is also proportional to the square of the scattering center velocity (Liu 2015). In realistic physical environments, two acceleration mechanisms are possibly present in the shock. A dominant mechanism depends on the ratio of acceleration rate (e.g., Jones 1994; Schlickeiser 2002; Protheroe \textit{et al}. 2004; Petrosian  2012).

{\bf\emph{Injection}}~~This is a typical acceleration process in which a fraction of high-energy thermal particles are selected from the background plasma easily to trigger a \emph{Fermi}-type acceleration. The injection process create a seed population of the non-thermal particle. This pre-acceleration process should be based on a resonance between low-energy particles and the turbulence waves, such as a few thermal particles in the downstream region being energized by an interaction with turbulence. These can induce the particles to multiply recross the shock and trigger a DSA process (Kirk \textit{et al.} 1992; Malkov \textit{et al}. 2001; Rieger \textit{et al}. 2007; Verkhoglyadova \textit{et al}. 2015).

{\bf\emph{Maximum Energies}}~~Generally, the plasma in astrophysics is highly conducting and fully ionized. As a result, no electric fields exist. However, in a collisionless plasma system, the particle acceleration could be realized by the electric field induced by the time-varying magnetic field. In this scenario, the \emph{Fermi} acceleration mechanism is an ultimately electrodynamic accelerator (Kirk \textit{et al}. 1992; Dermer \textit{et al}. 2009). The maximum energy is determined by one of two conditions. Either the shock runs out of time to accelerate particles or it runs out of space. Regarding time, the rate of acceleration is an important, but not solely determining, factor. An infinite planar shock that exists indefinitely will accelerate particles to arbitrarily large energies regardless of how quickly it accelerates them, and its steady-state solution will be a power law that goes on indefinitely. If one limits the size of the shock, the upper energy cutoff will appear where the mean free path of particles exceeds the scale size of the shock. If one limits the lifetime of the shock, the upper energy cutoff will be approximately the acceleration rate realized by the duration of the shock. In all cases, both of these are limiting factors, but usually one is the dominant limiter on the acceleration process.

{\bf\emph{Particle-Transport Equation}}~~We assume that a cold-background plasma is characterized by a electric field $E_{0}=0$ and large-scale magnetic field along with the\textit{ z} axis, $\vec{B_{0}}=B_{0}\vec{e}_{z}$ (Schlickeiser 1989; Achatz \textit{et al}. 1991). It is supposed that small-amplitude and stochastic plasma turbulence with $\delta\vec{E}\ll\delta\vec{B}\ll\vec{B_{0}}$, and $\langle\delta\vec{E}\rangle=\langle\delta\vec{B}\rangle=0$ exists in the rest frame of the plasma turbulence supporting fluid. The unperturbed background medium moves with a bulk speed $\vec{U}=U(z)\vec{e}_{z}$ and it is parallel to $\vec{B}_{0}$. The wave vectors of the magneto-hydrodynamic (MHD) turbulence waves areparallel or antiparallel to $\vec{B_{0}}$. Since $\delta\vec{E}$ is perpendicular to their corresponding wave vector, the waves can be regarded as transverse waves in which the shortest growth time in this condition (Tademaru 1969) with a dispersion relation $\omega^{2}=V_{A}^{2}k^{2}$ in the fluid frame exists (Petrosian \textit{et al}. 2004; Dermer \textit{et al}. 2009; Schlickeiser 2015), where $V_{A}\ll c$ is the Alfven velocity and $k$ is wave number. We consider the test-particle approximation case in which the gyro-radii of particles are smaller than the spatial plasma scales. These particles should not influence the plasma conditions. Thus, we can neglect the magnetic field amplification by the particle-streaming instabilities and shock structure fluctuation by the energetic particle pressure. The phase-space density of these particles is denoted by the function $f(\vec{X}, p, \mu, \phi, t)$, where $\vec{X}=(X, Y, Z)$ is the guiding center coordinates of energetic particles, and the spherical coordinates ($p, \mu, \phi$) are used in momentum space (Achatz \textit{et al}. 1991). In such a collisionless plasma, the particles are accelerated in the ambient electric field (Jokipii 2001). Although the gyro-radius of the energetic particle is larger than other microscopic lengths, it is usually small in macroscopic terms. In this view, the distribution function should be significantly independent of the gyration phase. We can then average over this coordinate. Introducing the general relativistic Vlasov equation (Webb 1985; Kirk \textit{et al}. 1987), we can deduce the Larmor-phase averaged Fokker-Planck transport equation in the quasi-linear approximation, in which the time and space coordinates are in the laboratory frame and the momentum coordinates are in the rest frame of the streaming plasma (e.g., Webb 1985; Kirk \textit{et al}. 1988; Schlickeiser 2015):
\begin{eqnarray}
\Gamma\biggl[1+\frac{Uv\mu}{c^{2}}\biggr]\frac{\partial f}{\partial t}&+&\Gamma[U+v\mu]\frac{\partial f}{\partial z}+\frac{\nu(1-\mu^{2})}{2L}\frac{\partial f}{\partial \mu}\nonumber\\&-&\alpha(z)\biggl(\mu+\frac{U}{v}\biggr)\biggl[\mu p\frac{\partial f}{\partial p}+(1-\mu^{2})\frac{\partial f}{\partial \mu}\biggr]\nonumber\\&=&\frac{\partial}{\partial\mu}\biggl[D_{\mu\mu}\frac{\partial f}{\partial \mu}+D_{\mu\sigma}\frac{\partial f}{\partial y_{\sigma}}\biggr]\nonumber\\&+&\frac{1}{p^{2}}\frac{\partial}{\partial y_{\omega}}p^{2}\biggl[D_{\omega\mu}\frac{\partial f}{\partial\mu}+D_{\omega\sigma}\frac{\partial f}{\partial y_{\sigma}}\biggr]\nonumber\\&+&S(\vec{X}, p, t)\;,
\label{Eq:1}
\end{eqnarray}
where $\Gamma=[1-U^{2}/c^{2}]^{-1/2}$ is the bulk Lorentz factor of the background plasma, $y_{\omega}$, $y_{\sigma}\in[p, X, Y]$ represent the three phase space variables with non-vanishing stochastic fields $\delta\vec{E}$ and $\delta\vec{E}$, $z=Z$, $S(\vec{X}, p, t)$ accounts for additional source of particles, $c$ is the speed of light, $v$ is the particle speed, $\alpha(z)=\Gamma^{3}dU(z)/dz$ is the rate of adiabatic deceleration or acceleration in relativistic flows (Schlickeiser 2015), and $L=-B_{0}(z)[dB_{0}(z)/dz]^{-1}$ is the focusing length (Roelof 1969). Because the mirror force in the large-scale inhomogeneous guiding magnetic field $B_{0}$ is proportional to $dB_{0}(z)/dz$ (Somov 2013), the third term on the left-hand side of Eq.(\ref{Eq:1}) represents the effect of the magnetic field convergence ($L<0$) or divergence ($L>0$). However, this term should vanish if the background magnetic field is uniform. $\Gamma[1+Uv\mu/c^{2}]$ and $\Gamma[U+v\mu]$ on the left-hand side of Eq. (\ref{Eq:1}) is relative to the Lorentz boost from the fluid frame to laboratory frame (Kirk \textit{et al.} 1988). When the background medium moves with a varying bulk speed ($\alpha(z)\neq0$), the fourth term on the left-hand side of Eq. (\ref{Eq:1}) are the connection coefficients of a general relativistic transformation between the laboratory frame and the mixed co-moving frame (Schlickeiser 2015). The terms on the right-hand side generally containing 16 different Fokker-Planck coefficients ($D_{\mu\mu}, D_{\mu\sigma}, D_{\omega\mu}, D_{\omega\sigma}$) describe the interaction of particles with Alfven waves supported by the background fluid. The Fokker-Planck coefficients were calculated by Steinacker \textit{et al}. (1992), Schlickeiser (2002), Petrosian \textit{et al}. (2004), and Dermer \textit{et al}. (2009). These coefficients depend on turbulent fields, such as magneto-static fields (vanishing turbulent electric fields), iso-spectral fields, and slab turbulence, and only $D_{\mu\mu}$ is non-zero and represents pitch-angle scattering (Blandford \textit{et al}. 1987; Schlickeiser 2015).	
		
In many cases, the gyro-radius of particles is much smaller than the scale of the spatial variation of the field in magnetized plasma. The particles are tied to the magnetic field lines (Petrosian  2012). As a result, the distributions are weakly variable in $X$ and $Y$. Therefore, we consider the distribution $f(z, p, \mu, t)$, which depends on the spatial coordinate $z$ along the field lines, the momentum $p$, the pitch angle or its cosine $\mu$, and the time. Then, Eq. (\ref{Eq:1}) can be rewritten as
\begin{eqnarray}
\Gamma\biggl[1+\frac{Uv\mu}{c^{2}}\biggr]\frac{\partial f}{\partial t}&+&\Gamma[U+v\mu]\frac{\partial f}{\partial z}+\frac{\nu(1-\mu^{2})}{2L}\frac{\partial f}{\partial \mu}\nonumber\\&-&\alpha(z)\biggl(\mu+\frac{U}{v}\biggr)\biggl[\mu p\frac{\partial f}{\partial p}+(1-\mu^{2})\frac{\partial f}{\partial \mu}\biggr]\nonumber\\&=&\frac{\partial}{\partial\mu}\biggl[D_{\mu\mu}\frac{\partial f}{\partial \mu}+D_{\mu p}\frac{\partial f}{\partial p}\biggr]\nonumber\\&+&\frac{1}{p^{2}}\frac{\partial}{\partial p}p^{2}\biggl[D_{p\mu}\frac{\partial f}{\partial\mu}+D_{pp}\frac{\partial f}{\partial p}\biggr]\nonumber\\&+&S(\vec{X}, p, t)\;.
\label{Eq:2}
\end{eqnarray}

Since the magnetic field component is much larger than the electric field component ($|\delta \vec{B}|=(c/V_{A})|\delta\vec{E}|$, $V_{A}\ll c$), the scattering by low-frequency MHD turbulence is so rapid that the pitch-angle scattering time $\tau_{sc}\sim 1/D_{\mu\mu}$ is shorter than other timescales ($\tau_{diff}\sim p^{2}/D_{pp}$, $\tau_{cross}\sim R/v$, loss timescale, acceleration timescale, etc., where \textit{R} is the size of the interaction region) and the distribution is close to isotropic in the rest frame of the moving background plasma (Blandford\textit{ et al}. 1987; Schlickeiser 1989; Petrosian 2012). In this case, we can use a diffusion approximation to energetic particles (Jokipii 1966), in which we define the anisotropic $g(z, p, \mu, t)$ and isotropic $F(z, p, t)$ parts of the distribution function by
\begin{equation}
f(z, p, \mu, t)=\frac{1}{2}F(z, p, t)+g(z, p, \mu, t)\,,
\label{Eq:3}
\end{equation}
with $g\ll f$ and $g=O(U/v)F$. Here, $F(z, p, t)=\int_{-1}^{+1}d\mu f(z, p, \mu, t)$ and $\int_{-1}^{+1}d\mu g(z, p, \mu, t)=0$. Inserting Eq. (\ref{Eq:3}) into Eq. (\ref{Eq:2}), if we neglect the terms that are higher than the second order of $(U/v)$,  we obtain the basic particle-transport equation for the ultra-relativistic particles in the non-relativistic flow with $\Gamma\sim1$, $V_{A}\ll v$, and $U\ll c$ (e.g., Kirk \textit{et al}. 1988; Schlickeiser 1989),
\begin{eqnarray}
\frac{\partial F}{\partial t}-S_{0}&=&\frac{\partial}{\partial z}\biggl(\kappa_{zz}\frac{\partial F}{\partial z}\biggl)+\frac{1}{p^{2}}\frac{\partial}{\partial p}\biggl(p^{2}a_{2}\frac{\partial F}{\partial p}\biggr)\nonumber\\&+&\frac{v}{4}\frac{\partial}{\partial z}(a_{1}\frac{\partial F}{\partial p})-\frac{1}{4p^{2}}\frac{\partial}{\partial p}\biggl(p^{2}va_{1}\frac{\partial F}{\partial z}\biggr)\nonumber\\&-&U\frac{\partial F}{\partial z}+\frac{p}{3}\frac{\partial U}{\partial z}\frac{\partial F}{\partial p}-\frac{\kappa_{zz}}{L}\frac{\partial F}{\partial z}\;,
\label{Eq:4}
\end{eqnarray}
where, $\kappa_{zz}=(v^{2}/8)\int^{+1}_{-1}[(1-\mu^{2})^{2}/D_{\mu\mu}]d\mu$ is the spatial diffusion coefficient, $a_{1}=\int^{+1}_{-1}[(1-\mu^{2})D_{\mu p}/D_{\mu\mu}]d\mu$ is the rate of adiabatic deceleration, $a_{2}=(1/2)\int^{+1}_{-1}[D_{pp}-(D_{\mu p}^{2}/D_{\mu\mu})]d\mu$ is the momentum diffusion coefficient, and $S_{0}=\langle S(z, p, t)\rangle$ is the source term. The coefficients $\kappa_{zz}$, $a_{1}$, and $a_{2}$ depend on the power spectrum of the magnetic field inhomogeneities. In the pioneering paper of Dung and Schlickeiser (1990a; 1990b), it is shown that the transport coefficient {\bf is} very sensitive to the chosen magnetic and cross-helicities of the Alfvenic turbulence.

Noting that
\begin{eqnarray}
\frac{v}{4}\frac{\partial}{\partial z}(a_{1}\frac{\partial F}{\partial p})&-&\frac{1}{4p^{2}}\frac{\partial}{\partial p}\biggl(p^{2}va_{1}\frac{\partial F}{\partial z}\biggr)=\frac{v}{4}\frac{\partial a_{1}}{\partial z}\frac{\partial F}{\partial p}\nonumber\\&-&\frac{1}{4p^{2}}\frac{\partial}{\partial p}\biggl(p^{2}va_{1}\biggr)\frac{\partial F}{\partial z}\,,
\label{Eq:5}
\end{eqnarray}
and $L\rightarrow\infty$ for a homogeneous background magnetic field, We rewrite Eq. (\ref{Eq:4}) as
\begin{eqnarray}
\frac{\partial F}{\partial t}&-&S_{0}=\frac{\partial}{\partial z}\biggl(\kappa_{zz}\frac{\partial F}{\partial z}\biggl)-\biggl[U+\frac{1}{4p^{2}}\frac{\partial}{\partial p}\biggl(p^{2}va_{1}\biggr)\biggr]\frac{\partial F}{\partial z}\nonumber\\&+&\frac{1}{p^{2}}\frac{\partial}{\partial p}\biggl(p^{2}a_{2}\frac{\partial F}{\partial p}\biggr)+\biggl(\frac{p}{3}\frac{\partial U}{\partial z}+\frac{v}{4}\frac{\partial a_{1}}{\partial z}\biggr)\frac{\partial F}{\partial p}\;.
\label{Eq:6}
\end{eqnarray}
Equation (\ref{Eq:6}) contains on the right-hand side terms representing spatial diffusion, spatial convection, momentum diffusion, and adiabatic deceleration or acceleration (convection). We note that the processes of spatial diffusion and convection transport the particles, as well as that the processes of momentum diffusion and convection accelerate the particles. It is well known that Fermi (1949; 1954) refers to the momentum diffusion and convection terms as \emph{Fermi} acceleration of second and first order, respectively.

We consider the case of the Alfven wave propagating forward in a homogeneous background magnetic field. In this case,  the Alfven wave propagating in only one direction does not give rise to momentum diffusion; that is, $a_{1}=4pV_{A}/(3v)$ and $a_{2}=0$ (Schlickeiser 1989; Dung and Schlickeiser 1990b). We obtain $1/(4p^{2})\partial (p^{2}va_{1})/\partial p=V_{A}$. If we assume a homogeneous turbulence, we find $\partial a_{1}/\partial z=0$ . Then, Eq. (\ref{Eq:6}) reduces to a standard particle-transport equation (Parker equation) for the diffusive shock acceleration (e.g., Parker 1965; Blandford and Ostriker 1978; Drury 1983):
\begin{equation}
\frac{\partial F}{\partial t}+U\frac{\partial F}{\partial z}=\frac{\partial}{\partial z}(\kappa_{zz}\frac{\partial F}{\partial z})+\frac{p}{3}\frac{\partial U}{\partial z}\frac{\partial F}{\partial p}+Q\;,
\label{Eq:7}
\end{equation}
where we assume $V_{A}\ll U$ and use the approximation with $U+V_{A}\approx U$.

We now consider the case in which the Alfven wave propagates with same intensity parallel and anti-parallel in both polarization states in a homogeneous background magnetic field. In this case, there is no adiabatic deceleration; that is, $a_{1}=0$. Noting that
\begin{eqnarray}
U\frac{\partial F}{\partial z}-\frac{p}{3}\frac{\partial U}{\partial z}\frac{\partial F}{\partial p}=\frac{\partial}{\partial z}\biggl(UF\biggr)-\frac{1}{p^{2}}\frac{\partial}{\partial p}\biggl(\frac{p^{3}}{3}F\frac{\partial U}{\partial z}\biggr)\,,
\label{Eq:8}
\end{eqnarray}
Eq. (\ref{Eq:4}) reduces to
\begin{eqnarray}
\frac{\partial F}{\partial t}-S_{0}&=&\frac{\partial}{\partial z}\biggl(\kappa_{zz}\frac{\partial F}{\partial z}-UF\biggr)\nonumber\\&+&\frac{1}{p^{2}}\frac{\partial}{\partial p}\biggl[p^{2}a_{2}\frac{\partial F}{\partial p}+\biggl(\frac{p}{3}\frac{\partial U}{\partial z}\biggr)p^{2}F\biggr]\;.
\label{Eq:9}
\end{eqnarray}
We argue that observations of AGN refer to volume-integrated intensities from a given source, so that for spatially constant acceleration rates one only needs the volume-integrated energy spectra of the radiating electrons. Generalizing to three spatial dimensions, the spatial operator in Eq. (\ref{Eq:9}) reduces to
\begin{eqnarray}
\frac{\partial F}{\partial t}-S_{0}&=&\nabla\cdot\biggl(\kappa_{zz}\nabla F-\vec{U}F\biggr)\nonumber\\&+&\frac{1}{p^{2}}\frac{\partial}{\partial p}\biggl[p^{2}a_{2}\frac{\partial F}{\partial p}+\biggl(\frac{p}{3}\nabla\cdot\vec{U}\biggr)p^{2}F\biggr]\;.
\label{Eq:10}
\end{eqnarray}
Adopting a spatially uniform acceleration region (e.g., Park \textit{et al}. 1995; Stawarz \textit{et al}. 2008; Mertsch 2011), and adding the terms of first-order \emph{Fermi} acceleration at shock waves $\dot{p}_{gain}$ and continuous losses $\dot{p}_{loss}$, the volume-integration of Eq. (\ref{Eq:10}) with Gauss' law (e.g., Schlickeiser \textit{et al}. 1987) provides that
\begin{eqnarray}
\frac{\partial n(p, t)}{\partial t}&=&\frac{\partial}{\partial p}\biggl[D_{pp}\frac{\partial n(p, t)}{\partial p}\biggr]\nonumber\\&-&\frac{\partial}{\partial p}\biggl[\biggl(\dot{p}_{gain}+\frac{2D_{pp}}{p}-\dot{p}_{loss}-\frac{p}{3}\nabla\cdot\vec{U}\biggr)n(p, t)\biggr]\nonumber\\&-&\frac{n(p, t)}{t_{esc}}+Q_{0}\,,
\label{Eq:11}
\end{eqnarray}
where $n(p, t)=\int_{V} dV[4\pi p^{2}F(p, \vec{X}, t)]$ denotes the volume-integrated electron spectrum, $D_{pp}=a_{2}$ is the momentum diffusion coefficient that denotes the interactive rate of the fluctuation electromagnetic field, $2D_{pp}/p$ denotes the second-order \emph{Fermi} acceleration rate, $(p/3)\nabla\cdot\vec{U}$ denotes the adiabatic deceleration losses, $n(p, t)/t_{esc}$ represents the rate at which particles leak out of the confinement region, and $Q_{0}=\int_{V} dV(4\pi p^{2}S_{0})$ denotes the volume-integrated population of injected electrons. Because the second-order \emph{Fermi} acceleration is an essential diffusion in momentum space, the diffusion term can broaden the injection spectrum and $2D_{pp}/p$ can result in a systematic shift to higher momenta (Mertsch 2011).

We further consider a magneto-static, iso-spectral, and slab turbulence for a relativistic flow. In this special case, the concise diffusion-convection equation is derived by Schlickeiser (2015):
\begin{eqnarray}
&&\Gamma\frac{\partial F}{\partial t}-\frac{\kappa_{zz}\Gamma}{L}\frac{\partial F}{\partial z}+\frac{\partial}{\partial z}\biggl[\Gamma\biggl(UF-\Gamma\kappa_{zz}\frac{\partial F}{\partial z}\biggr)\biggr]\nonumber\\&+&\frac{1}{p^{2}}\frac{\partial}{\partial p}\biggl(p^{2}\kappa_{pz}\Gamma\frac{\partial F}{\partial p}-\frac{\alpha(z) p^{3}F}{3}\biggr)=Q\;,
\label{Eq:12}
\end{eqnarray}
where $\kappa_{pz}=\frac{\alpha(z)vp}{12}(K +\frac{12U}{v^{3}}\kappa_{zz})$, and
$K=\int_{ - 1}^1 {\frac{{\mu {{\left( {1 - {\mu ^2}} \right)}^2}}}{{{D_{\mu \mu }}\left( \mu  \right)}}}d\mu$.
In the non-relativistic limit with $\Gamma  \approx 1$, the diffusion-convection equation becomes (Schlickeiser 2015)
\begin{eqnarray}
&&\frac{{\partial F}}{{\partial t}} - \frac{{{\kappa _{zz}}}}{L}\frac{{\partial F}}{{\partial z}} + \frac{\partial }{{\partial z}}\left( {UF - {\kappa _{zz}}\frac{{\partial F}}{{\partial z}}} \right) \nonumber\\&+&\ \frac{1}{{{p^2}}}\frac{\partial }{{\partial p}}\left( {{p^2}{\kappa _{pz}}\frac{{\partial F}}{{\partial p}} - \frac{{\partial U}}{{\partial z}}\frac{{{p^3}F}}{3}} \right) = Q\;.
\label{Eq:13}
\end{eqnarray}
We note that the term of containing $\kappa _{pz}$ is absent from the transport equation (\ref{Eq:7}) for the diffusive shock acceleration, but appears in Eqs. (\ref{Eq:12}) and (\ref{Eq:13}). This new term results from the connection coefficients in Eq. (\ref{Eq:1}). If we assume $\frac{{{\partial ^2}F}}{{\partial {z^2}}}\sim\frac{{{\partial ^2}F}}{{\partial p\partial z}}\sim\frac{{\partial F}}{{\partial z}}$, since $\frac{1}{{{p^2}}}\frac{\partial }{{\partial p}}\left( {{p^2}{\kappa _{pz}}\frac{{\partial F}}{{\partial z}}}\right) \sim v\frac{{\partial F}}{{\partial z}}K+pv\left( {1 + \frac{U}{v}} \right)K\frac{{{\partial ^2}F}}{{\partial z\partial p}}$ and $\frac{\partial }{{\partial z}}\left( {{\kappa _{zz}}\frac{{\partial F}}{{\partial z}}} \right) \sim {v^2}K\frac{{{\partial ^2}F}}{{\partial {z^2}}}$, we can deduce that $\frac{1}{{{p^2}}}\frac{\partial }{{\partial p}}\left( {{p^2}{\kappa _{pz}}\frac{{\partial F}}{{\partial p}}} \right) \ll \frac{\partial }{{\partial z}}\left( {{\kappa _{zz}}\frac{{\partial F}}{{\partial z}}} \right)$. This suggests that the new term can be neglected when we deduce Eq. (\ref{Eq:4}) from Eq. (\ref{Eq:2}). Because the correction to the standard spectral index $M$ [see further discussions in case (9) in Section 3.4] in the relativistic energy regime is small in normal circumstances (Schlickeiser 2015), it is also reasonable for neglecting this term in Eq. (\ref{Eq:7}).

However, due to the highly anisotropic of distribution function $f$ for an ultra-relativistic flow, we should give up the diffusion approximation and return to the full transport Eq. (\ref{Eq:2}) that considers a angular dependence to deal with the relativistic shock (Kirk\textit{ et al}. 1992). When we consider pitch-angle diffusion, neglect injection, and assume a uniform background magnetic field, the one-dimensional steady-state transport equation for the ultra-relativistic particle can be written as
\begin{eqnarray}
&&\Gamma \left(\mu+\frac{U}{c}\right)\left\{c\frac{\partial f}{\partial z}-\frac{dU}{dz}\Gamma^2\left[\mu p\frac{\partial f}{\partial p}+\left({1-\mu^2}\right)\frac{\partial f}{\partial\mu} \right]\right\}\nonumber\\&&=\frac{\partial}{\partial\mu }\left(D_{\mu\mu}\frac{\partial f}{\partial \mu} \right)\;,
\label{Eq:14}
\end{eqnarray}
where we use a approximation $v\approx c$. We consider a parallel shock and the distance from the shock front $z$ is measured in the shock frame. In order to simplify Eq. (\ref{Eq:14}), we assume $D_{\mu\mu}$ with the form $D_{\mu\mu}=D_{1}\left(\mu\right)D_{2}\left(p,z\right)$, where $D_2\left(p,z\right)$ is a function depending on $p$ and $z$. We define a variable $\tau=\int\limits_0^z {{D_2}} \left( {p,z'} \right)dz'\Gamma$ to replace $z$, and we can rewrite Eq. (\ref{Eq:14}) as (Keshet \textit{et al}. 2005; Schneider\textit{ et al}. 1989)
\begin{eqnarray}
\left(\mu+\frac{U}{c}\right)\frac{\partial f(\mu, p, \tau)}{\partial\tau}=\frac{\partial}{\partial\mu}D_{1}(\mu)\frac{\partial f(\mu, p, \tau)}{\partial \mu }\;,
\label{Eq:15}
\end{eqnarray}
where ${D_1}\propto\left( {1 - {\mu ^2}}\right)$ is independent of momentum $p$ for isotropic diffusion.
\section{Solution of the continuity equation}
We can gain insight into the nature of the coupled energetic particle transport in astrophysical plasma through the analytic solution of the diffusion equation. Many investigators tempt to solve the equation analytically (e.g., Schlickeiser 1984; 1985; Park \textit{et al}. 1995; Becker \textit{et al.} 2006; Stawarz \textit{et al.} 2008; Mertsch 2011). In the following, we compile the solutions from the literature.
\subsection{Volume-integrated transport equation}
We consider an isotropic power-law MHD turbulent wave spectrum $W(k)\propto k^{-q}~(1\leq q \leq 2,~k_{1} \leq k \leq k_{2})$. Here, $W(k)dk$ represents the energy density between the wave number $k$ and $k+dk$, which is small compared with the unperturbed magnetic field energy density $B_{0}^{2}/(8\pi)$. The spectral indexes $q=2$, $q=5/3$, $q = 3/2$, and $q=1$ describe the hard-sphere approximation, Kolmogorov case, Kraichnan case, and Bohm limit, respectively.

In the case of an ltra-relativistic electron with rest mass ${m_e}$, the particle momentum can be approximated by the particle Lorentz factor $p=\gamma$, where the particle energy and momentum are expressed in units of $m_{e}c^{2}\ $and $m_{e}c$, respectively. Therefore, we rewrite Eq. (\ref{Eq:11}) with $D_{\gamma\gamma}=D(\gamma)$, and $A(\gamma)=\dot{\gamma}_{gain}+2D_{\gamma\gamma}/\gamma$ in the form
\begin{eqnarray}
\frac{\partial n(\gamma, t)}{\partial t} &=& \frac{\partial}{\partial\gamma}\left[D(\gamma)\frac{\partial n(\gamma, t)}{\partial\gamma}\right]-\frac{\partial}{\partial\gamma}\left[ A(\gamma)-\dot{\gamma}_{loss}\right]n(\gamma, t)\nonumber\\&-&\frac{n(\gamma, t)}{t_{esc}} + Q_{0}\;,
\label{Eq:16}
\end{eqnarray}
where, we neglect the adiabatic deceleration losses, and the advective coefficient $\dot{\gamma}_{loss}=4\sigma_{T}(U_{B}+U_{rad})\gamma^{2}/(3m_{e}c)=\eta\gamma^{2}$ describes both the synchrotron and IC cooling with the magnetic energy density $U_{B}$, radiation field energy density $U_{rad}$, and Tomson cross-section $\sigma_{T}$. The momentum diffusion coefficient is $D(\gamma)=\gamma^{2}/t_{D}\propto \gamma^{q}$; here, the characteristic acceleration timescale $t_{D}\approx c^{-1}\zeta^{-1}\beta_{A}^{-2}\lambda_{2}^{(q-1)}(m_{e}c^{2})^{(2-q)}(eB_{0})^{(q-2)}\gamma^{(2-q)}$ describes stochastic particle-wave interactions, where $\lambda_{2}=2\pi/k_{1}$ is the maximum wavelength of the Alfven modes, and $\beta_{A}=V_{A}/c$, and $\xi=(\delta B)^{2}/B_{0}^{2}\leq 1$ denotes the ratio between the turbulence energy density and unperturbed magnetic field energy density. Here, we assume the escape timescale $t_{esc} \approx3L^{2}\zeta c^{-1}\lambda_{2}^{(1-q)}(m_{e}c^{2})^{(q-2)}(eB_{0})^{(2-q)}\gamma^{(q-2)}$ (Stawarz \textit{et al}. 2008). We define the energy-loss timescale $t_{loss}= 1/(\eta\gamma)$, and systemic energy gain timescale $t_{A}=\gamma/A(\gamma)=\gamma/[2\gamma/t_D+\gamma/t_{sh}]\approx[2/t_{D}+U^{2}/(4\kappa_{zz})]^{-1}$, where $t_{sh}=3dz/dU\approx3(U_{1}-U_{2})^{-1}(\kappa_{1}/U_{1}- \kappa_{2}/U_{2})\approx4\kappa_{zz}/U^{2}$ is the timescale of the first-order \emph{Fermi} acceleration at a shock front with non-relativistic speed $U$. $U_{1}(U_{2})$ and $\kappa_{1}( \kappa_{2})$ are the flow velocities and spacial diffusion coefficients upstream (downstream) in the shock frame, respectively (Schlickeiser 1984; 1985). We assume the spatial diffusion coefficients $\kappa_{1}$, $\kappa_{2}$, and $\kappa_{zz}$ are independent of the particle energy $\gamma$ ; see Eq. (\ref{Eq:15}).

For convenience, we introduce the energy variable $x=\gamma/\gamma_{0}$, where $\gamma_{0}$ is some chosen value of the particle energy, such as the energy of injected particles. We define the dimensionless time $\tau=t/t_{D}$. In these scenarios, the particle energy spectrum, the source, the stochastic acceleration, and escape timescales can be written as $N=\gamma_{0}n$, $Q_{1}=\tau_{D}\gamma_{0}Q_{0}$, $t_{D}\approx\tau_{D}\gamma^{(2-q)}$ with $\tau_{D}=c^{-1}\zeta^{-1}\beta_{A}^{-2}\lambda_{2}^{(q-1)}(m_{e}c^{2})^{(2-q)}(eB_{0})^{(q-2)}$, and $t_{esc}\approx\tau_{esc}\gamma^{(q-2)}$ with $\tau_{esc}=3L^{2}\zeta c^{-1}\lambda_{2}^{(1-q)}(m_{e}c^{2})^{(q-2)}(eB_{0})^{(2-q)}$, respectively (Stawarz \textit{et al.} 2008). In these cases, Eq. (\ref{Eq:16}) could be written as
\begin{eqnarray}
\frac{\partial N}{\partial\tau}&=&\frac{\partial}{\partial x}\left[x^{q}\frac{\partial N}{\partial x}\right]-\frac{\partial}{\partial x}\left[\left(2x^{(q-1)}+ax\right)N\right] \nonumber\\&-&\varepsilon x^{(2-q)}N + Q_{1}\;,
\label{Eq:17}
\end{eqnarray}
where $a=U^{2}\tau_{D}/(4\kappa_{zz})-\theta_{x}$, $\theta_{x}=\tau_{D}/t_{loss}$, and $\epsilon=\tau_{D}/\tau_{esc}$. The steady-state equation (\ref{Eq:17}) is expressed asw
\begin{eqnarray}
\frac{\partial}{\partial x}\left[x^{q}\frac{\partial N}{\partial x}\right]&-&\frac{\partial}{\partial x}\left[\left(2x^{(q-1)}+ax\right)N\right]\nonumber\\&-&\varepsilon x^{(2-q)}N + Q_{1}=0\;.
\label{Eq:18}
\end{eqnarray}
Owing to the finite wave-number ranges with $k_{1}\leq k\leq k_{2}$, the particle energy is in a finite range $x_{1}\leq x\leq x_{2}$ with $0<x_{1}$ and $x_{2}<\infty$. These result in a regular boundary value problem (Stawarz \textit{et al}. 2008). Park and Petrosian (1995) discussed the singular ($x \in \left[ {0,\infty } \right]\ $) boundary condition in detail. In this special case, there is a steady-state solution if there is a non-zero particle flux through the $x_{1}$ but no particle flux at $x_{2}$. The steady-state solution implies the conservation of particle number that is generated  by a kind of competition between the injection and escape of the particles or energy losses (Park and Petrosian 1995).

When the mono-energetic particle populations with energy $x_{0}$ are injected at time ${\tau _0}$, we can obtain $Q_{1}\propto\delta(x-x_{0})\delta(\tau-\tau_{0})$. Then, we can expect a Green's-function solution of Eq. (\ref{Eq:17}).  Generally, we can deduce the Green's-function solution using the Sturm-Liouville eigenfunction expansion theory for second-order differential equations. The process of solving Eq. (\ref{Eq:17}) can be summarized as the following: (1) we transform the equation into the self-adjoint form, where we replace $\partial/\partial t$ by eigenvalues, and then obtain the eigenfunction of the self-adjoint operator in the imposing boundary conditions for the flux; (2) according to the completeness relation, the Green's function can be expanded an infinite sum of the eigenfunctions. Solving the equation with the initial condition, we can obtain the expanding coefficients. Then, the particular solution associated with an arbitrary source distribution can be computed using the integral convolution of the Green's function and the source (Schlickeiser 1984; 1985; Park \textit{et al}. 1995; Becker \textit{et al}. 2006).

The program of solving the steady-state equation (\ref{Eq:18}) is simpler (Stawarz \textit{et al}. 2008). The Green's-function solution in the steady state should be approximated by a power law or an exponential function with one or several critical energies, such as the injection energy ${x_0}$, or the equilibrium energy that is determined by $t_{D}=t_{esc}$ when the systematic energy loss is absence; otherwise, $t_{A} = t_{loss}$. If all of timescales satisfy the same function form, the preferred energy scale can not be formed. Therefore, the steady-state solution can be approximated by a power law; otherwise, there is a critical energy that results in competition among escape, cooling losses, and acceleration (Park and Petrosian 1995).
\subsection{ Special cases}
		
\begin{figure}
  \centering
  \includegraphics[width=9 cm]{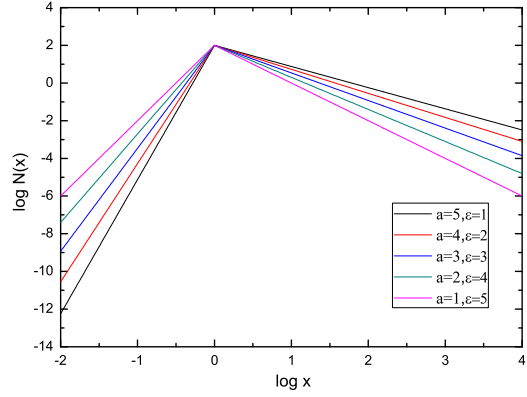}
  \caption{EED as given in Eq. (\ref{Eq:19}) at different $a$ and $\epsilon$ with $x_{0}=1$.}\label{fig:1}
\end{figure}

{\bf\emph{Case (1): Acceleration, escape, and mono-energetic injection}}~~In this case, we do not take into account the energy loss, i.e., ${\theta _x} = 0$. We assume that $q=2$ and continuous injection $Q_{1} \propto \delta \left( {x - {x_0}} \right)$. In this scenario, the Green's-function solution of Eq. (\ref{Eq:18}) is (Park \textit{et al}. 1995; Becker \textit{et al}. 2006)
\begin{equation}
N(x)\propto\left\{
\begin{array}{ll}
x^{(1 + a)/2} + \sqrt{(3+a)^{2}/4+\varepsilon}~~~(x < x_{0}),\\
x^{(1 + a)/2} - \sqrt{(3+a)^{2}/4+\varepsilon}~~~(x > x_{0}).
\end{array} \right.
\label{Eq:19}
\end{equation}
We show the EED as given in Eq. {\ref{Eq:19}} at different $a$ and $\epsilon$ in Figure {\ref{fig:1}}.

\begin{figure}
  \centering
  \includegraphics[width=9 cm]{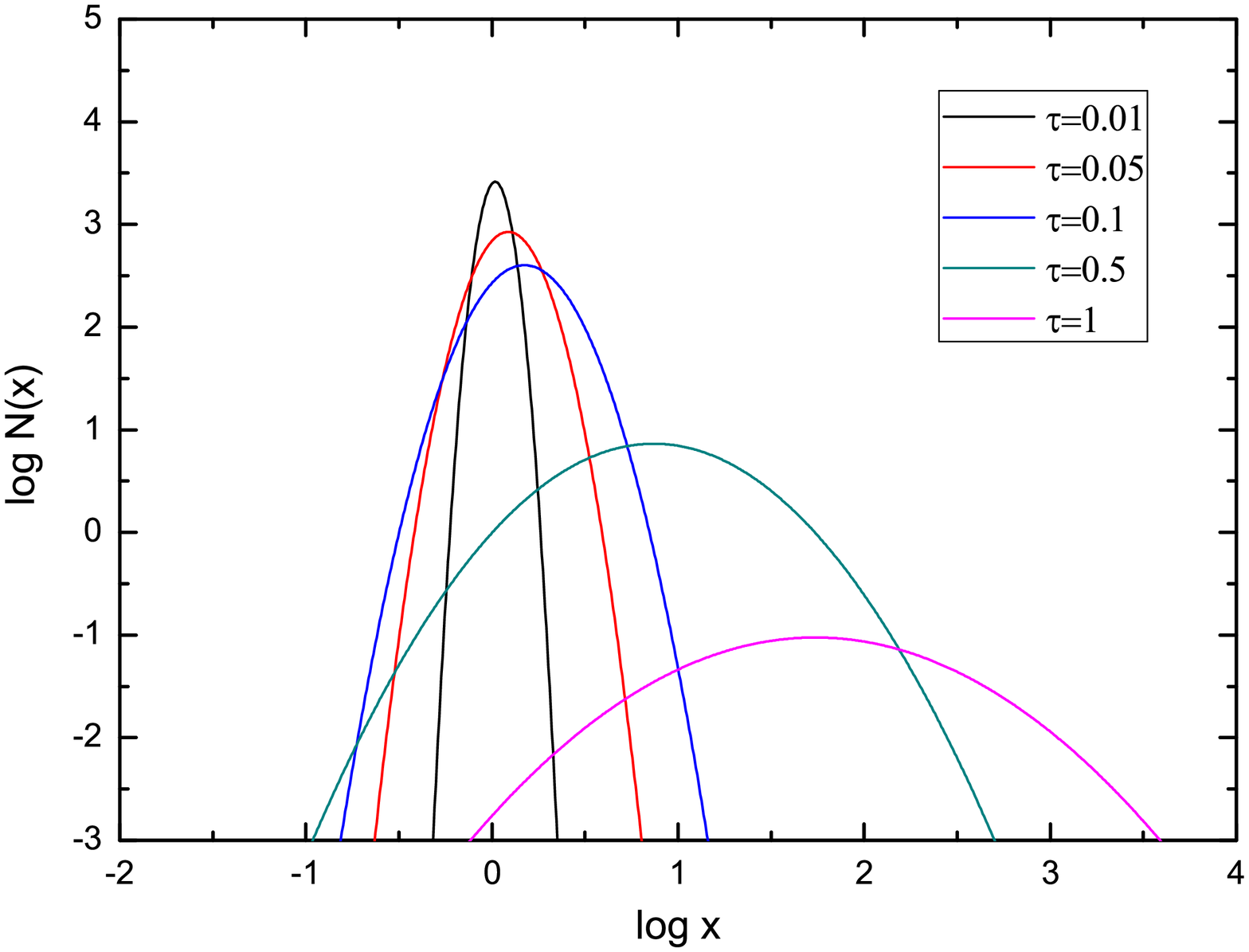}\\
  \caption{Time-dependent EED as given in Eq.  {\ref{Eq:20}} with $x_{0}=1$, $a=3$, and $\epsilon=3$.}\label{fig:2}
\end{figure}

Since both the escape timescale and the acceleration timescale are independent of the particle energy in this case, the spectrum shows a broken power law at a critical energy $x_{0}$. However, if we assume $q<2$, the stationary solution shows a power-law spectrum in the energy ranges $x \le {x_0}$, and a quasi-exponential cutoff in the energy ranges $x>x_{0}$ (Becker \textit{et al.} 2006). This spectrum indicates the fact that an efficient particle acceleration dominates the escape in lower energy ranges for any $q$, while in higher energy ranges the escape timescale decreases with increasing particle energy, and the escape results give a curve spectrum for $q<2$. Furthermore, in an instantaneous mono-energetic injection $Q_{1}\propto\delta(\tau  )\delta(x-x_{0})$, the time-dependent solution displays a log-parabolic distribution (Park \textit{et al}. 1995; Becker \textit{et al}. 2006):

\begin{equation}
N(x, \tau)\propto\frac{e^{-\varepsilon\tau}}{x\sqrt{4\pi\tau}}\exp \left\{-\frac{[\ln(x/x_{0})-(a+3)\tau]^{2}}{4\tau} \right\}\;.
\label{Eq:20}
\end{equation}
In Figure {\ref{fig:2}}, we show the time-dependent EED as given in Eq.  (\ref{Eq:20}).

\begin{figure}
  \centering
  \includegraphics[width=9 cm]{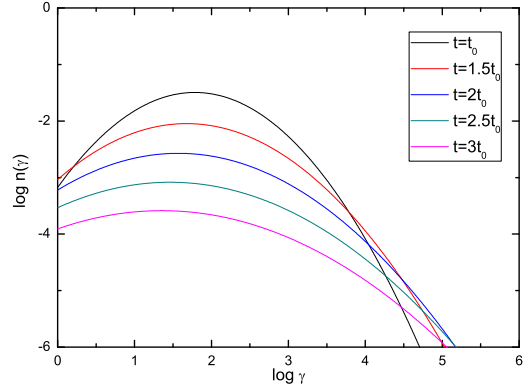}\\
  \caption{EED as given in Eq. {\ref{Eq:21}} at different $t$ with $t_{0}=6.0\times10^{-7}$~s.}\label{fig:3}
\end{figure}

{\bf \emph{Case (2): Time-dependent stochastic acceleration, and mono-energetic and instantaneous injection}}~~In this case, we do not take into account the energy loss, i.e., ${\theta_x} = 0$. We assume $q=2$ and an instantaneous mono-energetic injection $Q_{0} = {n_0}\delta (t)\delta (x - {x_0})$, where $n_{0}$ is the particle number density. Using an integral transform, we can find the time-dependent solution of Eq. (\ref{Eq:16}) ( Kardashev 1962; Paggi 2010),
\begin{equation}\
n(\gamma, t) = N_{0}(\gamma/\gamma_{0})^{-s_{1}-r_{1}\log(\gamma/\gamma_{0})},
\label{Eq:21}
\end{equation}
where $N_{0}=n_{0}\exp[-(A_{1}+A_{2})^{2}/4A_{2}]/2\sqrt{\pi A_1}\gamma_{0}$, $A_{1}=\int_{t_{0}}^{t}1/t_{A}(t)dt'$, $A_{2}=\int_{t_{0}}^{t}1/t_{D}(t)dt'$, $s_{1}=(1-A_{1}/A_{2})/2$, and the curvature $r_{1}=\ln(10)/4A_{2}$. This is a log-parabolic spectrum. We show the EED as given in Eq. (\ref{Eq:21}) at different $t$ in Figure {\ref{fig:3}}.  It can been seen that the spectral curvature is produced by the momentum-diffusion process. The spectrum can also be obtained to use an approach of intuitive statistical description (Massaro \textit{et al}. 2004; Tramacere \textit{et al}. 2011).	

\begin{figure}
  \centering
  \includegraphics[width=9 cm]{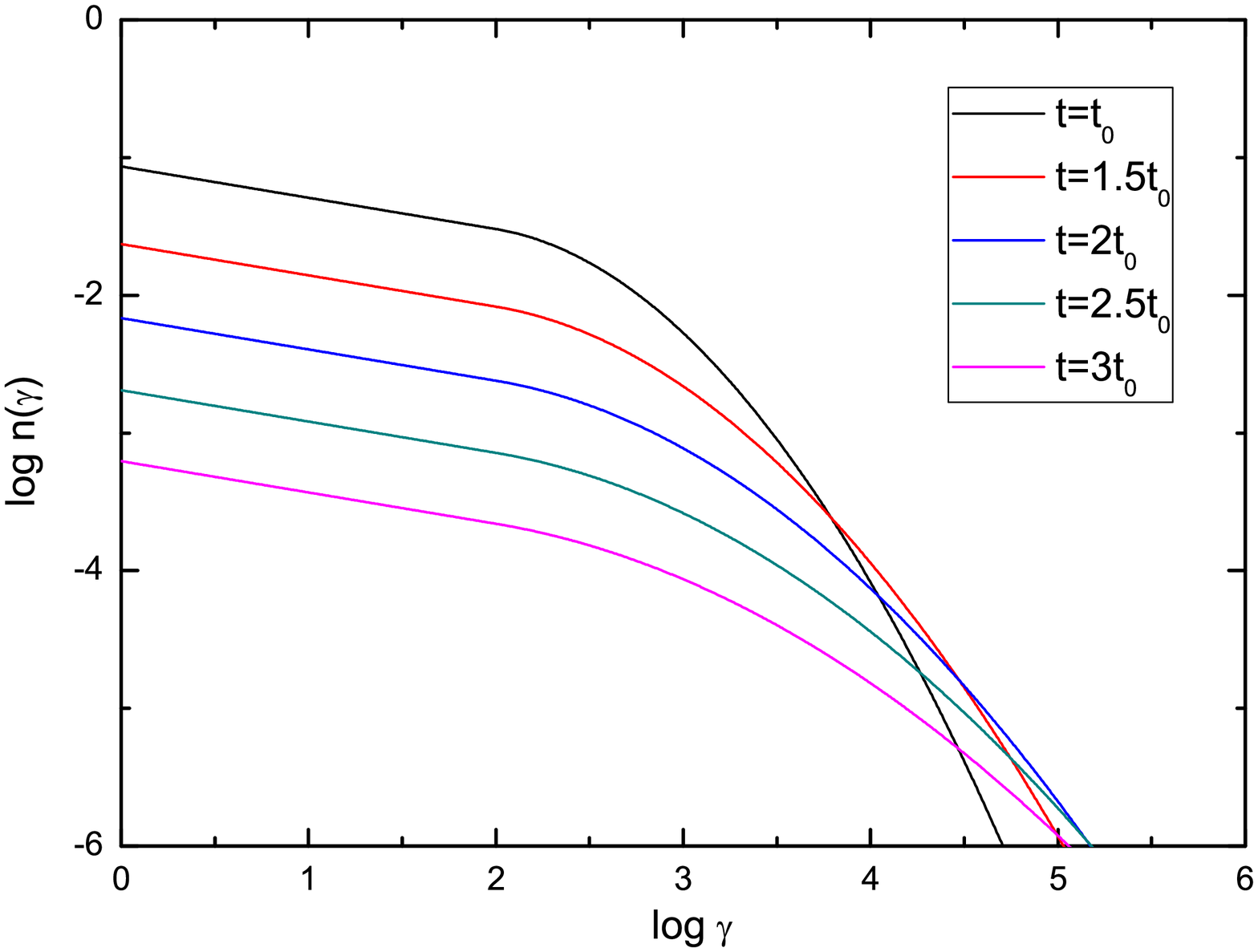}\\
  \caption{EED as given in Eq. {\ref{Eq:22}} at different $t$ with $t_{0}=6.0\times10^{-7}$~s for $s_{2}=s_{1}$ and $r_{2}=r_{1}$.}\label{fig:4}
\end{figure}

\begin{figure}
  \centering
  \includegraphics[width=9 cm]{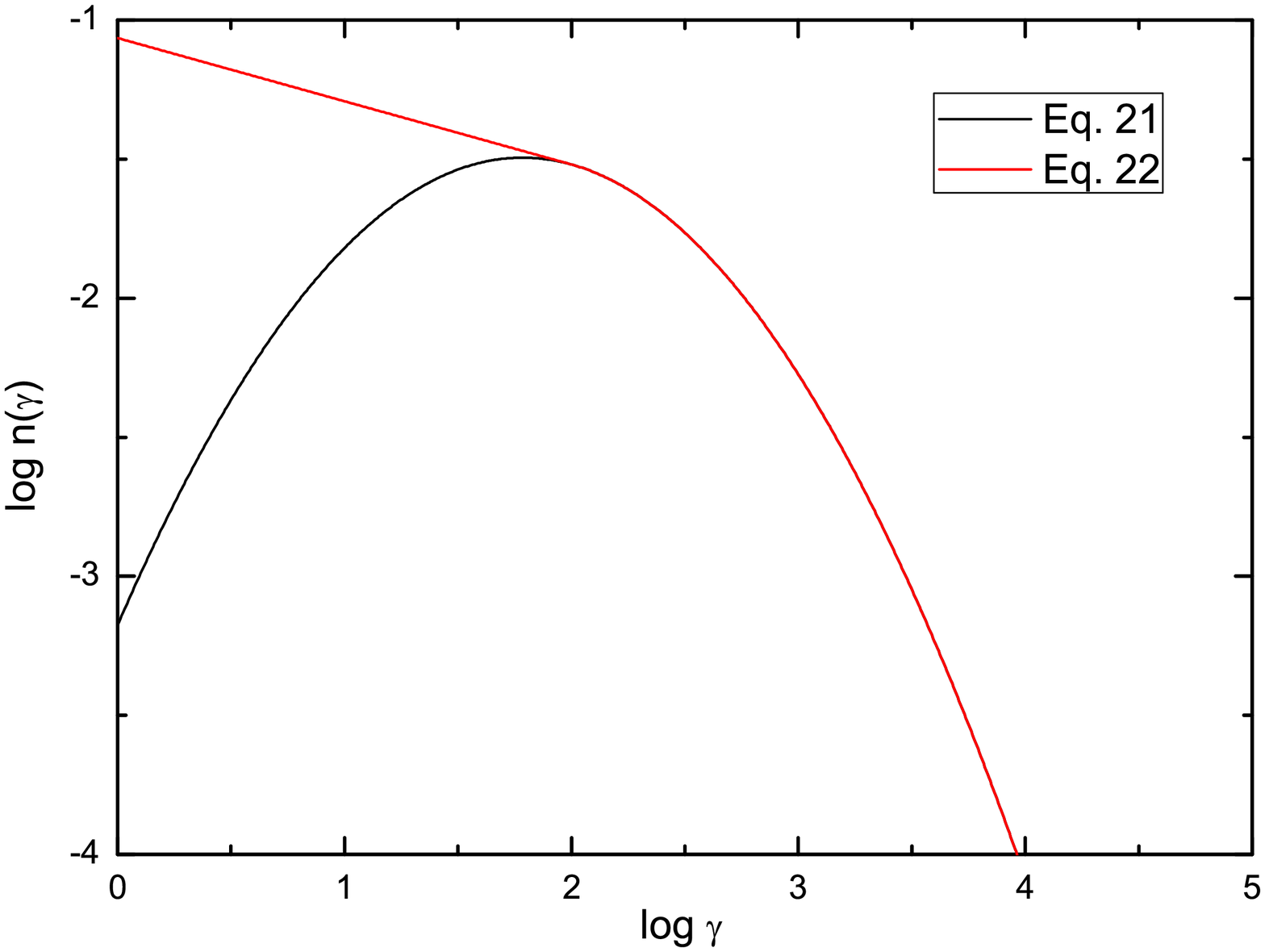}\\
  \caption{Comparison between the EED given by Eq. {\ref{Eq:21}} and that given by Eq. {\ref{Eq:22}}.}\label{fig:5}
\end{figure}

In addition, if we assume that the acceleration probability is a constant for low-energy particles, we can find (Massaro et al. 2006)
\begin{equation}
n(\gamma, t)\propto \left\{
\begin{array}{ll}
(\gamma/\gamma_{0})^{-s_{2}}\;,~~~~~\gamma \leq \gamma_{0},\\
(\gamma/\gamma_{0})^{-s_{2}-r_{2}\log(\gamma/\gamma _{0})}\;,~~~\gamma >\gamma_{0},
\end{array} \right.
\label{Eq:22}
\end{equation}
where $s_{2}=1-q_{1}/2-\log(g/\gamma_{0}^{q_{1}})/\log(\epsilon_{1})$, $r_{2}=q_{1}/2\log(\epsilon_{1})$, $\epsilon_{1}$ is the fraction of energy gain,  and $g$ and $q_{1}$ are positive constants that are relevant to the acceleration probability of electron. Since, without a mechanism that can compensate for the energy-gain process, we can obtain a time-dependent spectrum and the steady-state solution cannot be expected. We show the EED as given in Eq. {\ref{Eq:17}} at different $t$ in Figure {\ref{fig:4}}. In order to distinguish between Eqs. {\ref{Eq:16}} and (\ref{Eq:22}), we compare the EED given by Eq. {\ref{Eq:21}} and that given by Eq. (\ref{Eq:22}) in Figure {\ref{fig:5}}.

\begin{figure}
  \centering
  \includegraphics[width=9 cm]{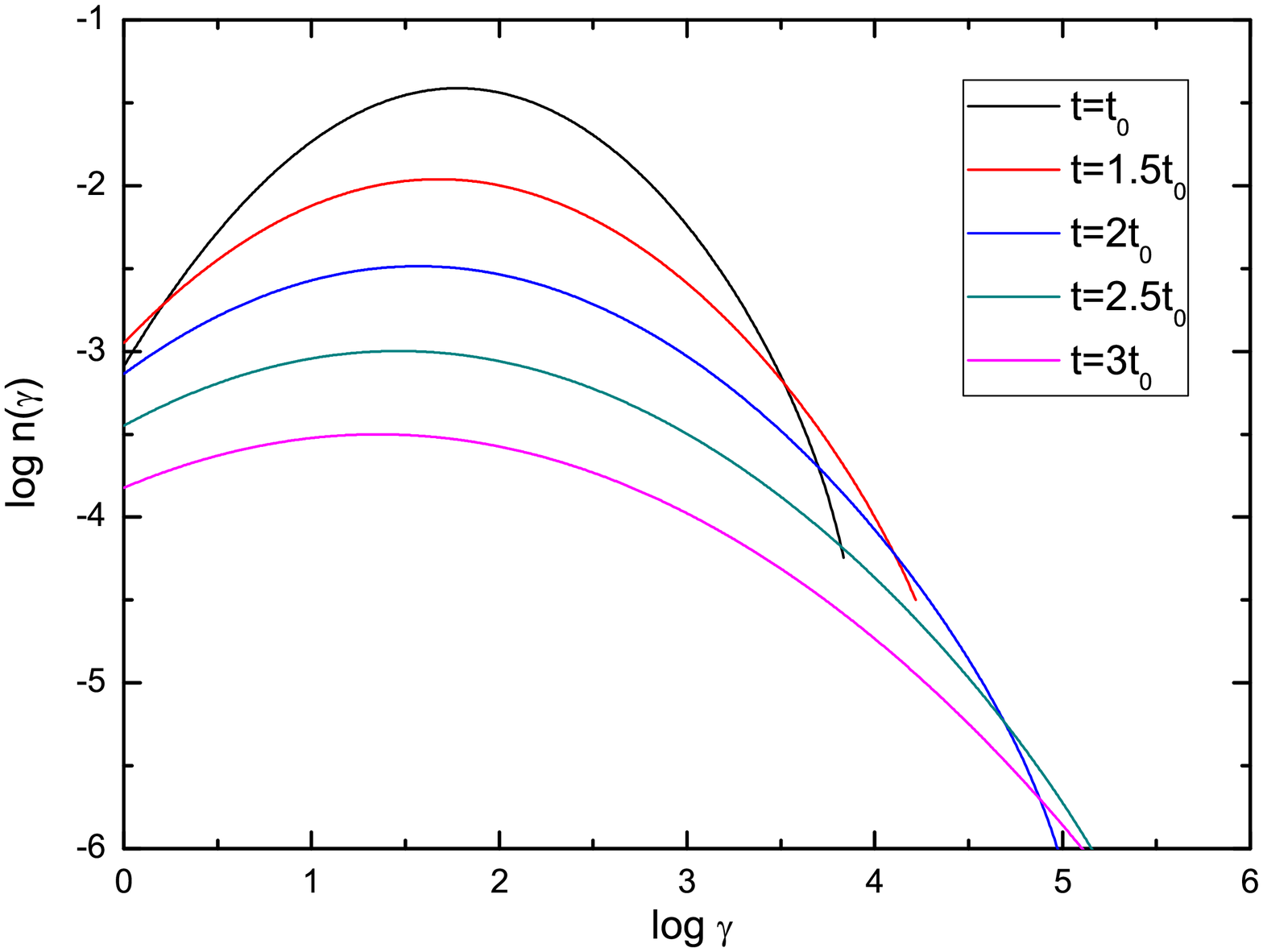}\\
  \caption{EED as given in Eq. (\ref{Eq:23}) at different $t$ with $t_{0}=6.0\times10^{-7}$~s.}\label{fig:6}
\end{figure}

{\bf\emph{Case (3): Stochastic acceleration, mono-energetic and instantaneous injection, and synchrotron or/and IC cooling}}~~In this case, we assume $q=2$ and an instantaneous mono-energetic injection $Q_{0}=n_{0}\delta(t)\delta(x-x_{0})$, where $n_{0}$ is the particle number density. In this scenario, the time-dependent solution of Eq. (\ref{Eq:11}) in the limit of ${\gamma _0}\eta {t_A} \ll 1$ is (Paggi 2010)
\begin{equation}\
n(\gamma, t)\approx N_{0}(\gamma/\gamma_{0})^{-s_{1}-r_{1}\log(\gamma/\gamma_{0})}\left[1+q_{1}(1-\gamma/\mathop\gamma\limits^{-} )\right]\;,
\label{Eq:23}
\end{equation}
where $\overline{\gamma}=\gamma_{0}10^{(3-2s_{1})/(4r_{1})}$ and $A_{1}=t/t_{A}$, $A_{2}=t/t_{D}$. We show the EED as given in Eq. (\ref{Eq:23}) at different $t$ in Figure {\ref{fig:6}}.

\begin{figure}[tb]
  \centering
  \includegraphics[width=9 cm]{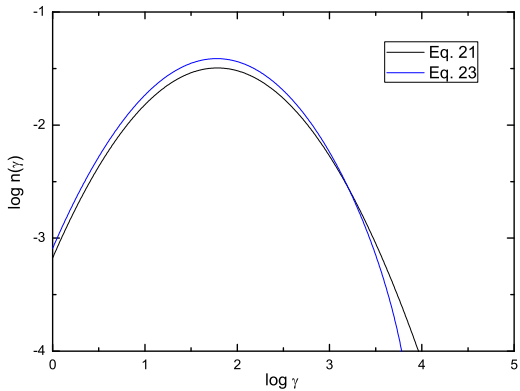}\\
  \caption{Comparison between the EED with radiative cooling (Eq. (\ref{Eq:23})) and without radiative cooling (Eq. (\ref{Eq:21}))}\label{fig:7}
\end{figure}

We compare  the EED with radiative cooling (Eq. (\ref{Eq:23})) and without radiative cooling (Eq. (\ref{Eq:21})) in Figure {\ref{fig:7}}. We find that the synchrotron and/or IC cooling change(s) the electron distribution at higher energy with a factor $q_{1}(1-\gamma/\overline\gamma)$. As shown in Figure {\ref{fig:7}}, the effect of radiative cooling
is to move high-energy electrons to lower energies, resulting in an increased curvature and a steepened electron distribution at high energies.

The radiative cooling can compensate for the energy-gain process,  and we can expect a stationary solution
\begin{eqnarray}
n(\gamma)\propto\gamma^{t_{D}/t_{A}}\exp[-\eta t_{D}(\gamma - 1)]\;.
\label{Eq:24}
\end{eqnarray}
When it satisfies $a_{1}=0$, the EED shows an ultra-relativistic Maxwellian distribution with $n(\gamma)\propto\gamma^{2}\exp[-\eta t_{D}(\gamma-1)]$ (e.g., Katarzynski \textit{et al}. 2006; Zheng and Zhang 2011a; 2011b).

\begin{figure}
  \centering
  \includegraphics[width=9 cm]{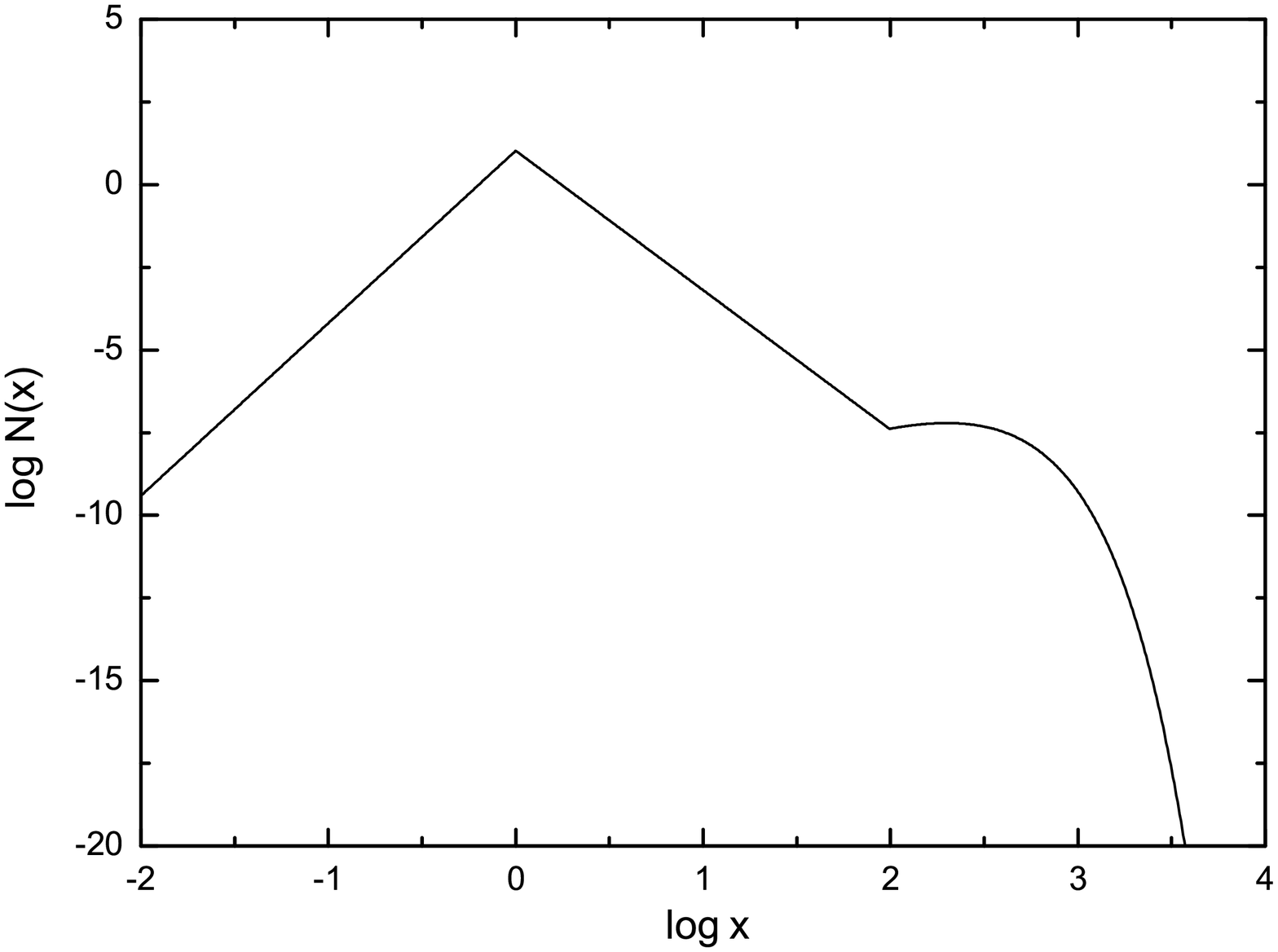}\\
  \caption{EED as given in Eq. (\ref{Eq:25}) with $x_{0}=1$, $x_{eq}=100$, and $\epsilon=20$.}\label{fig:8}
\end{figure}

{\bf\emph{Case (4): Acceleration, escape, mono-energetic injection, and synchrotron or /and IC cooling}}~~In this case, we assume $q=2$ and a mono-energetic injection $Q_{1}\propto\delta(x-x_{0})$. We define an equilibrium energy $x_{eq}$ by $t_{D}(x_{eq})=t_{loss}(x_{eq})$. We can approximate the solution of Eq. (\ref{Eq:18}) in the limit of $x_{1}<<x_{0}<<x_{eq}<<x_{2}$ (Stawarz\textit{ et al}. 2008),
\begin{eqnarray}
N(x )\propto\left\{
\begin{array}{lll}
\frac{1}{2\sigma+1}x_{0}^{-\sigma-2}x^{\sigma+1}\;,~~~x < x_{0} \\
\frac{1}{2\sigma+1}x_{0}^{-\sigma-2}x^{-\sigma}\;,~~~x_{0} < x << x_{eq}, \\
\frac{\Gamma (\sigma-1)}{\Gamma(2\sigma+2)}x_{0}^{-\sigma-1}x_{eq}^{-\sigma-2}x^{2}e^{-x/x_{eq}}\;~~~x_{eq} \le x \\
\end{array}\right.\;
\label{Eq:25}
\end{eqnarray}
where $\sigma=-1/2+\sqrt{9/4+\epsilon}$. We show the EED as given in Eq. (\ref{Eq:25}) in Figure {\ref{fig:8}}. It can be seen that the acceleration dominates the radiative loss, and the escape and acceleration timescales depend on the energy in the same form. We can approximate the spectrum by a broken power law in the energy range of $x_{1}<x< x_{eq}$. As for the higher energy range, the interplay between the loss and acceleration produce a pile-up bump $x^{2}\exp(-x/x_{eq})$ around the energy $x_{eq}$.

\begin{figure}
  \centering
  \includegraphics[width=9 cm]{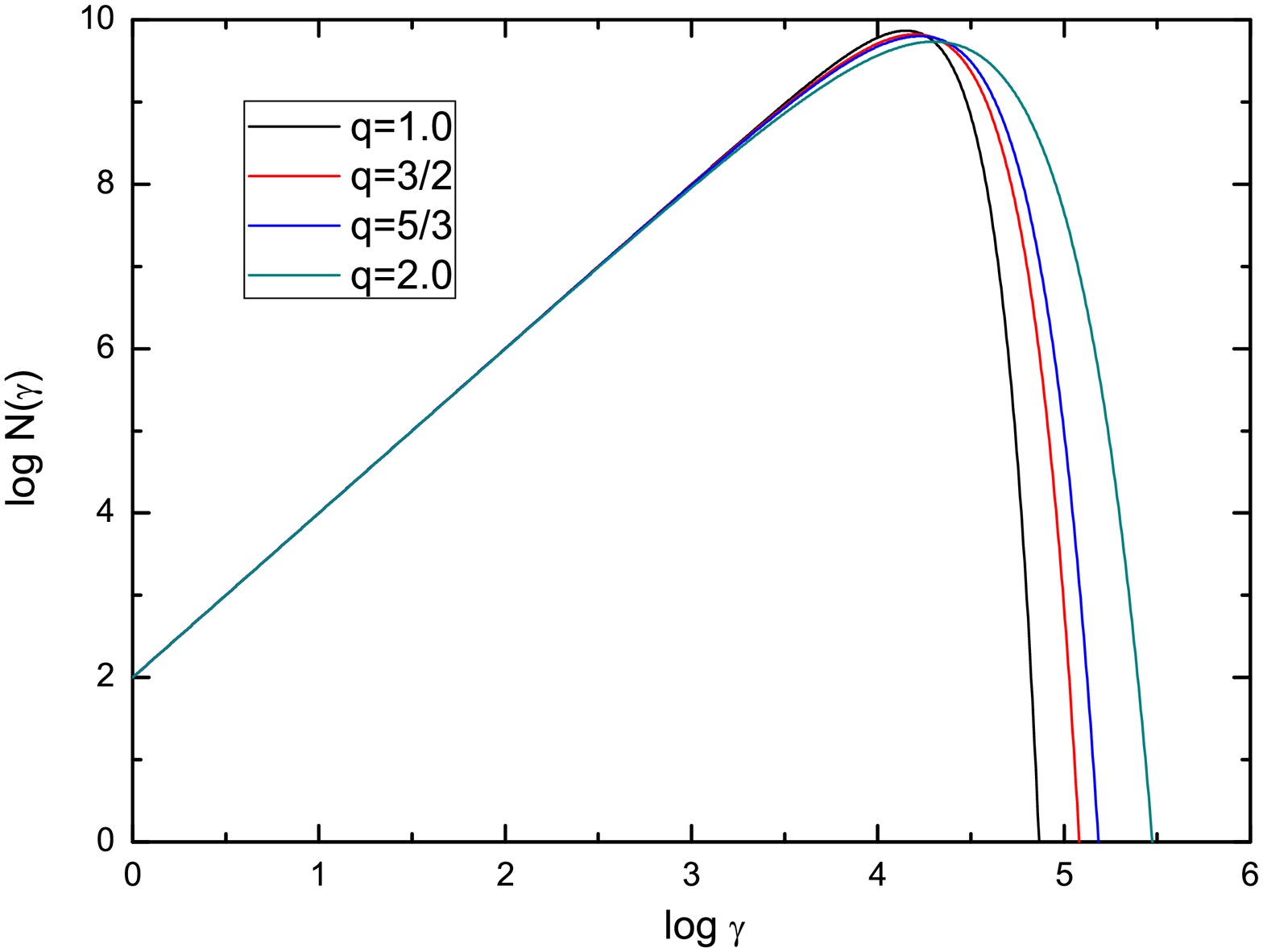}\\
  \caption{EED as given in Eq. (\ref{Eq:26}) at $q=1.0$, $q=3/2$, $q=5/3$, and $q=2.0$.}\label{fig:9}
\end{figure}

In more general cases, the solution shows a modified ultra-relativistic Maxwellian distribution at high energy if escape is inefficient ($t_{esc}\to\infty$) (Stawarz \textit{et al}. 2008),
\begin{equation}
N(\gamma ) \propto \gamma^{2}\exp \left[-(\gamma/\gamma_{eq})^{b}/b\right],
\label{Eq:26}
\end{equation}
where $b=3-q$, and $\gamma_{eq}$ is determined by $t_{D}(\gamma_{eq})=t_{loss}(\gamma_{eq})$. We show the EED as given in Eq. (\ref{Eq:26}) in Figure {\ref{fig:9}}.

\subsection{Negligible energy diffusion}
If we neglect the energy diffusion process - that is, $D(\gamma)=0$ - the Eq. (\ref{Eq:16}) can become a simple first-order differential equation,
\begin{equation}
\frac{\partial n(\gamma, t)}{\partial t}=-\frac{\partial}{\partial\gamma}\left(\frac{\gamma}{t_{A}}-\dot{\gamma}_{loss}\right)n(\gamma, t)-\frac{n(\gamma, t)}{t_{esc}}+Q_{0}\;,
\label{Eq:27}
\end{equation}
where we assume the escape timescale ${t_{esc}}$ is a constant, such as ${t_{esc}}=({\kappa _1}/{U_1}-{\kappa_2}/{U_2})/U_2$ in diffusive shock acceleration. This equation has been widely applied to model the emission of blazars (e.g., Kirk \textit{et al.} 1998; Chiaberge and Ghisellini 1999; B$\rm\ddot{o}$ttcher \textit{et al}. 2002; Finke and Becker 2014). The analytical solutions of Eq. (\ref{Eq:27}) in different conditions have been given in the past  (e.g. Kardashev 1962; Kirk \textit{et al}. 1998; Dermer \textit{et al}. 2009). In the following, based on the integral transform or the method of characteristics, we give the analytical solution of Eq. (\ref{Eq:27}) in the different cases, and then discuss its physical implications in some special cases.

{\bf\emph{Case (5): Radiative cooling and injection}}~~In this case, we assume that the electron population with a power-law electron spectrum $Q_{0}=n_{0}\gamma^{-m}~(\gamma_{1}\le\gamma\le\gamma_{2})$ is injected. We do not take into account the acceleration and escape [$1/t_A=0$ and $n(\gamma, t)/t_{esc}=0$].
We find the time-dependent solution of Eq. (\ref{Eq:27}) with an initial condition $n(\gamma,0)=0$ (Kardashev 1962),
\begin{eqnarray}
n(\gamma, t)&=&\frac{n_{0}\gamma^{-( m + 1)}}{\eta(\gamma-1)}\left[1 -(1 -\eta t\gamma)^{(\gamma-1)} \right]\ \nonumber\\&\approx&\left\{
\begin{array}{ll}
n_{0}t\gamma ^{-m}\;,~~~\gamma << \frac{1}{\eta t} \\
n_{0}\frac{\gamma ^{-(m + 1)}}{\eta(m-1)}\;,~~~\gamma>> \frac{1}{\eta t} \\
\end{array} \right.\;.
\label{Eq:28}
\end{eqnarray}
The physical {\bf implication} of this case is as follows: The electrons are accelerated at the shock front and then they form a power-law distribution with $n_{0}\gamma^{-m}$. These energetic electrons drift away from the shock front into the radiation zone in the downstream. Owing to the radiative cooling, the spectral index of emission electrons in the higher-energy regime is changed, while the spectral index of emission electrons in the low-energy regime remains inconsistent with that of the injection electrons.

\begin{figure}
  \centering
  \includegraphics[width=9 cm]{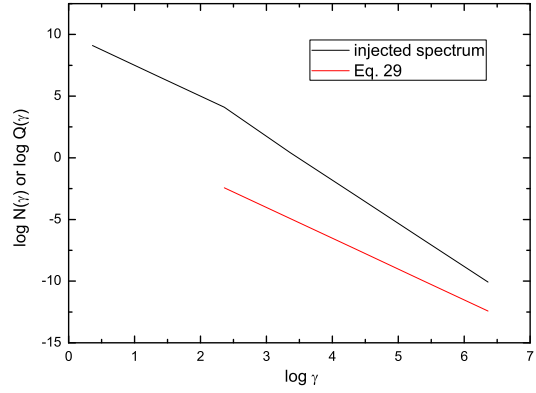}\\
  \caption{Injected spectrum and EED as given in Eq. (\ref{Eq:29}).}\label{fig:10}
\end{figure}

In light of the fact of that the particle flux of injection electrons can be balanced by the radiative cooling and escape, and that we adopt a no-particle-flux condition with $\dot{\gamma}n(\gamma, t)=0$ at $\gamma_{2}$, a steady-state solution of Eq. (\ref{Eq:27}) can be obtained:
\begin{eqnarray}
n(\gamma)&=&\frac{1}{\eta\gamma^{2}}\int_{\gamma}^{\gamma_{2}}Q_{1}(\gamma)d\gamma \nonumber\\&\simeq & \left\{
\begin{array}{ll}
\frac{n_{0}}{\eta}\gamma^{-m}\;,~~~\gamma \leq \gamma_{1}\\
\frac{n_{0}}{\eta}\gamma ^{-(m+1)}\;,~~~\gamma_{1} \leq \gamma<<\gamma_{2}
\end{array} \right.\;.
\label{Eq:29}
\end{eqnarray}
It can be seen that the spectral index of the steady-state spectrum is in agreement with the spectral index of the time-dependent spectrum. We show the EED as given in Eq. (\ref{Eq:29}) in Figure {\ref{fig:10}}.

\begin{figure}
  \centering
  \includegraphics[width=9 cm]{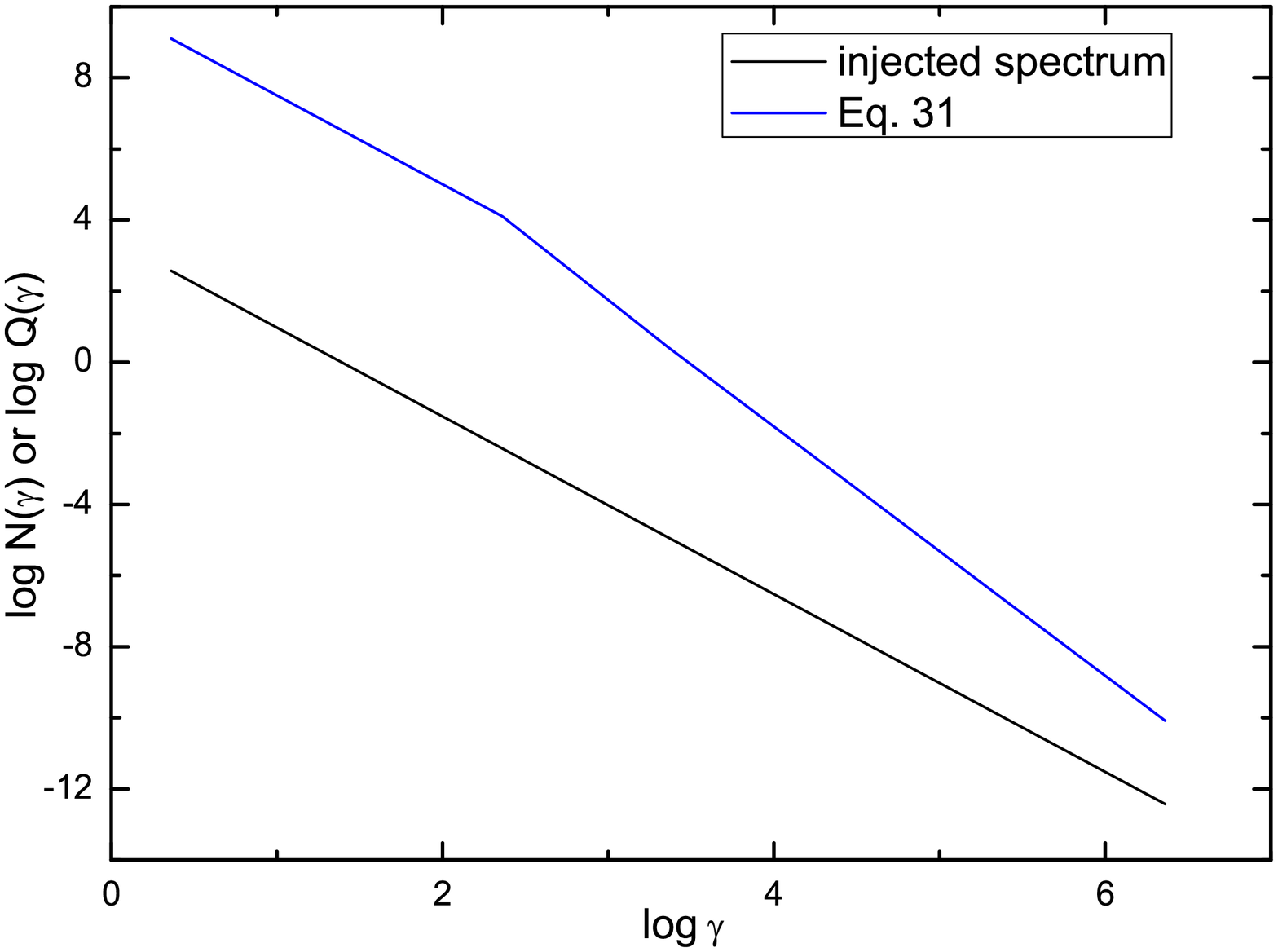}\\
  \caption{Injected spectrum and EED as given in Eq. {\ref{Eq:31}}.}\label{fig:11}
\end{figure}

{\bf\emph{Case (6): Radiative cooling, injection, and escape}}~~In this case, we assume that the electron population with a power-law electron spectrum $Q_{0}=n_{0}\gamma^{-m}~(\gamma_{1}\le\gamma\le\gamma_{2})$ is injected. We do not take into account the acceleration ($1/t_{A}=0$). In this scenario, there is a stationary solution of Eq. (\ref{Eq:27}) (Dermer \textit{et al}. 2009):
\begin{eqnarray}
n(\gamma)=\frac{1}{\eta\gamma^{2}}\int_{\gamma}^{\gamma_{2}}d\gamma'n_{0}\gamma'^{-m}\exp\left(-\int_{\gamma}^{\gamma'}\frac{d\gamma''}{t_{esc}\eta\gamma''} \right)\;.
\label{Eq:30}
\end{eqnarray}
When the injection electron energy satisfies $\gamma_{1}<\gamma_{c}$, the electrons show a slow cooling mode (e.g., Finke 2013; Zheng \textit{et al}. 2018). In this mode, we can approximate $n\left( \gamma  \right)$ as
\begin{equation}
n(\gamma, t) \propto \left\{
\begin{array}{ll}
\gamma^{-m}\;,~~~\gamma_{1}\leq\gamma\leq \gamma_{c}\\
\gamma^{-(m+1)}\;,~~~\gamma_{c}<\gamma << \gamma_{2}
\end{array} \right.\;.
\label{Eq:31}
\end{equation}
We show the EED as given in Eq. (\ref{Eq:31}) in Figure {\ref{fig:11}}.

\begin{figure}
  \centering
  \includegraphics[width=9 cm]{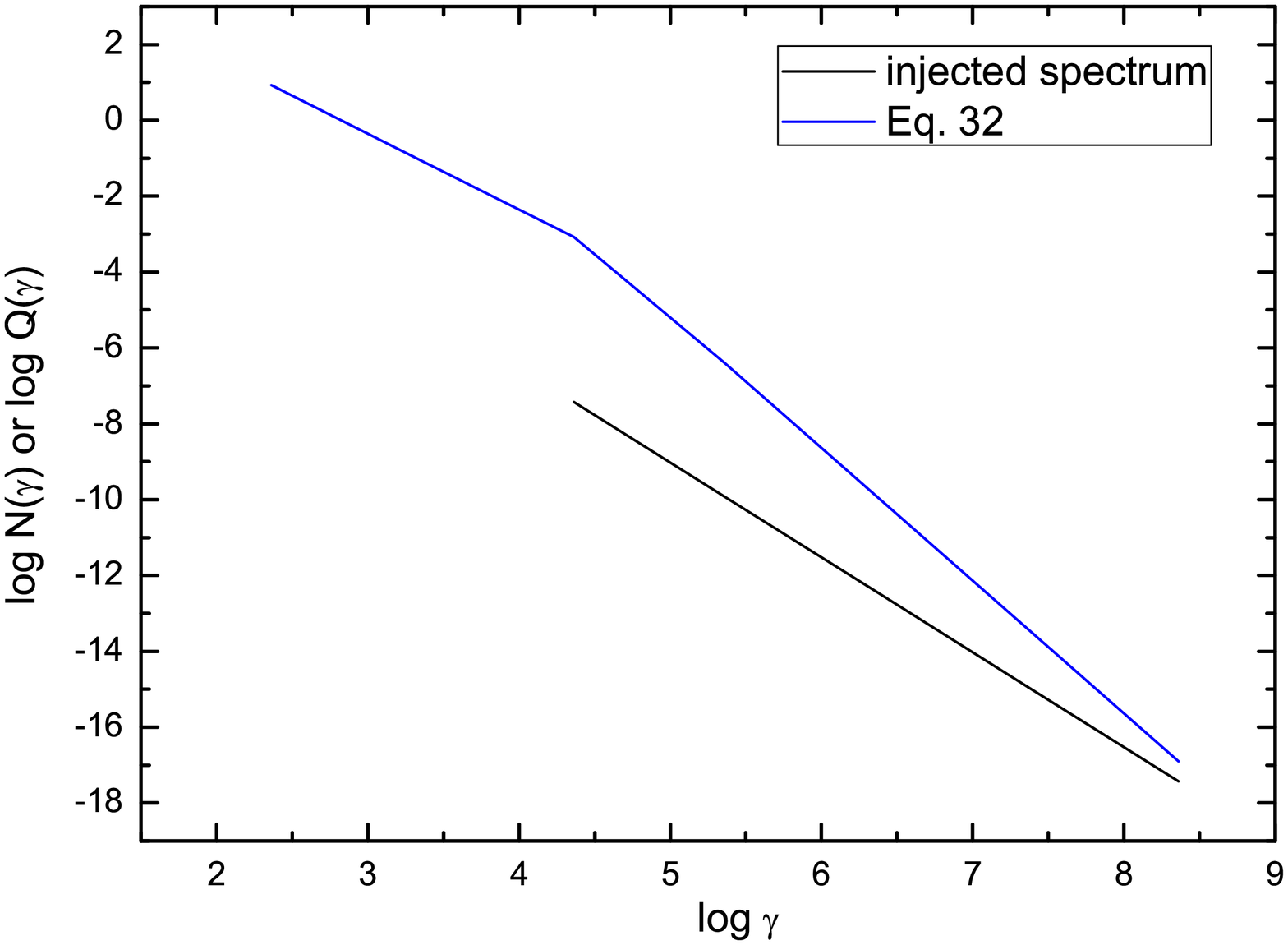}\\
  \caption{Injected spectrum and EED as given in Eq. (\ref{Eq:32}).}\label{fig:12}
\end{figure}

Otherwise, the electrons show a fast cooling mode. In this mode, we can approximate $n\left( \gamma  \right)$ as
\begin{eqnarray}
n(\gamma, t) \propto \left\{
\begin{array}{ll}
\gamma^{-2}\;,~~~\gamma_{c}\le\gamma\le\gamma_{1}\\
\gamma^{-(m+1)}\;,~~~\gamma_{1}<\gamma<<\gamma_{2}
\end{array} \right.\;,
\label{Eq:32}
\end{eqnarray}
where $\gamma_{c}=1/\eta t_{esc}$ is a critical energy that is deduced by $t_{esc}=t_{loss}(\gamma)$. We show the EED as given in Eq. (\ref{Eq:32}) in Figure {\ref{fig:12}}.

{\bf\emph{Case (7): Escape and systematic acceleration}}~~In this case, we do not take into account the particle injection. We assume that the escape scale is $t_{esc}\to\infty$. We argue that the turbulent acceleration can pick up a fraction of thermal particles from the background plasma to higher energy. Since the equilibrium can be generated by competition between the acceleration and escape of particles, where the acceleration is approximated as the injection, there is a stationary solution of Eq. (\ref{Eq:27}),
\begin{equation}
n(\gamma) \propto \gamma ^{-(t_{A}+t_{esc})/t_{esc}}\;.
\label{Eq:33}
\end{equation}
Equation (\ref{Eq:33}) shows a simple power-law spectrum. This spectrum is same as that of cases (5) and (6), in which it is assumed that electrons with a power-law distribution are injected into the radiative region and radiative cooling is taken into account.

\begin{figure}
  \centering
  \includegraphics[width=9 cm]{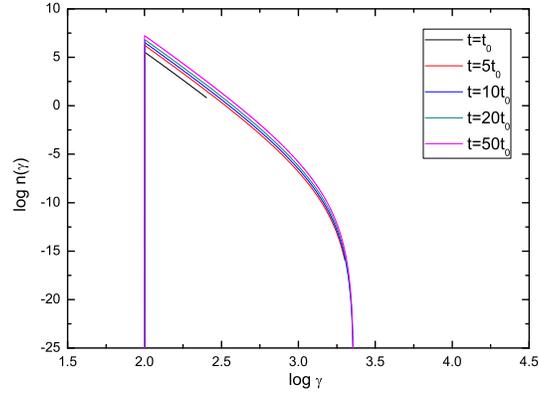}\\
  \caption{Time-dependent EED as given in Eq. (\ref{Eq:34}) at different $t$ with $t_{0}=10^{6}$~s, $t_{A}=t_{0}$, and $t_{esc}=0.1t_{0}$. To distinguish the spectrum at different times, we multiply different coefficients and $n(\gamma)$ together. }\label{fig:13}
\end{figure}

{\bf \emph{Case (8): Radiative cooling, mono-energetic injection, escape, and systematic acceleration}}~~In this general case, we assume that mono-energetic electron populations with $Q_{0}=n_{0}\delta(\gamma-\gamma_{0})~~( \gamma\le\gamma_{m}(t))$ are injected. In the initial condition $n(\gamma,0)=0$, the solution of Eq. (\ref{Eq:27}) can be approximated as a power-law distribution in the energy ranges $\gamma<<\gamma_{max}$ with a spectral index $M=1+t_{A}/t_{esc}$ (Kirk \textit{et al}. 1998),
\begin{eqnarray}
n(\gamma ,t)&=&\frac{d}{\gamma^2}(\frac{1}{\gamma}-\frac{1}{\gamma _{\max }})^{(t_{A}-t_{esc})/t_{esc}}\nonumber\\&\times&\Theta\left[\gamma_{m}(t)-\gamma\right]\Theta(\gamma-\gamma_0)\;,
\label{Eq:34}
\end{eqnarray}
where $d=n_{0}t_{A}\gamma_{0}^{t_{A}/t_{esc}}(1-\gamma_{0}/\gamma_{max})^{-t_{A}/t_{esc}}$, $\gamma_{m}(t)=[1/\gamma _{max}+(1/\gamma_{0}-1/\gamma_{max})e^{-t/t_{A}}]^{-1}$, the maximum energy $\gamma_{max}=1/(\eta t_{A})$ is derived from the relation $t_{loss}=t_{A}(\gamma)$, and $\Theta(x-x_{0})$ is the Heaviside function. We show the time-dependent EED as given in Eq. (\ref{Eq:34}) in Figure {\ref{fig:13}}.

\begin{figure}
  \centering
  \includegraphics[width=9 cm]{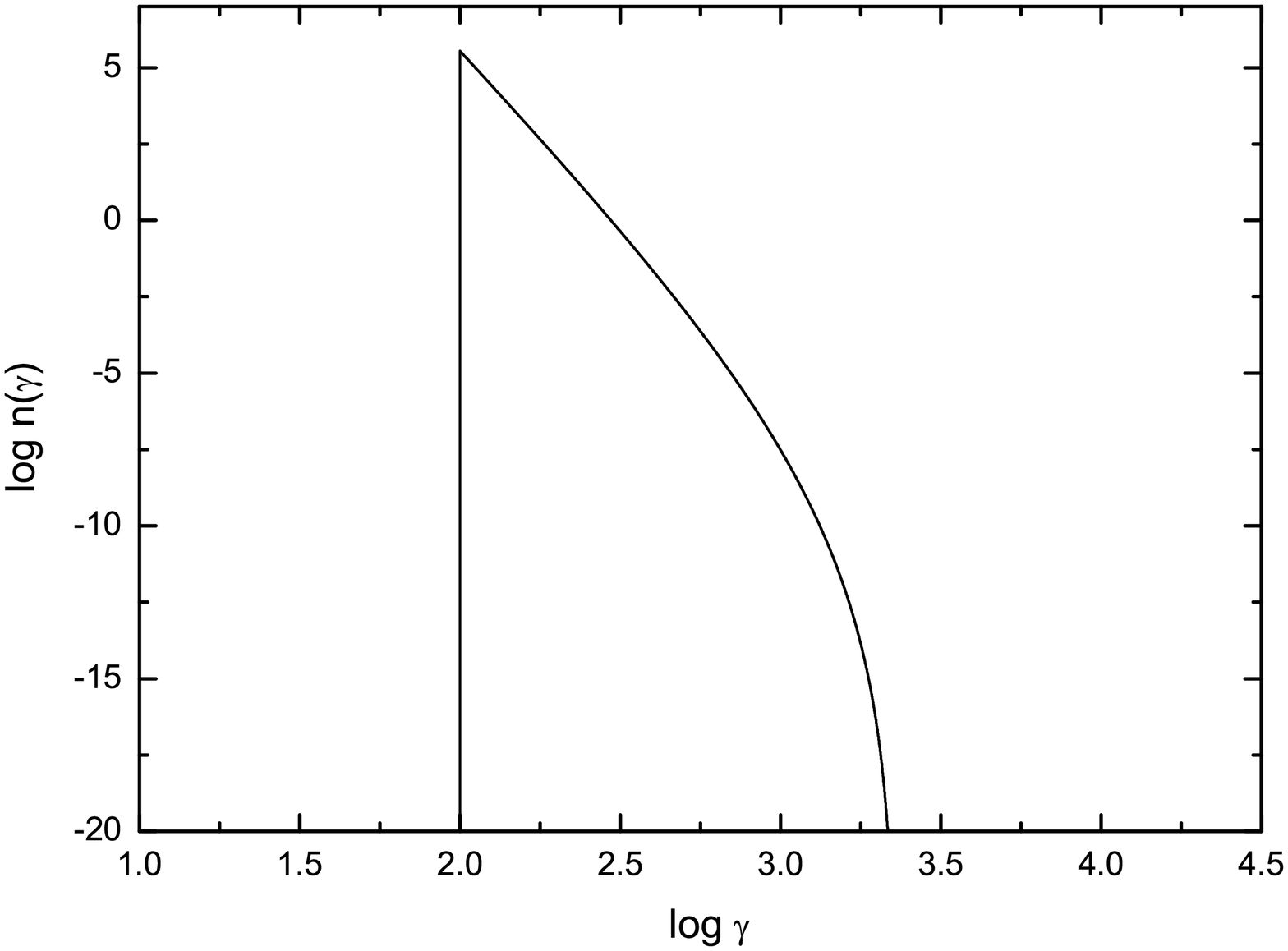}\\
  \caption{Steady-state EED as given in Eq. (\ref{Eq:35}).}\label{fig:14}
\end{figure}

When $t \to \infty$ is satisfied, we obtain the steady-state spectrum:
\begin{eqnarray}
n(\gamma)&=&\frac{d}{\gamma^{2}}(\frac{1}{\gamma}-\frac{1}{\gamma_{\max}})^{(t_{A}-t_{esc})/t_{esc}}\nonumber\\&\times&\Theta(\gamma_{\max}-\gamma)\Theta(\gamma-{\gamma_0}).
 \label{Eq:35}
\end{eqnarray}
We show the steady-state EED as given in Eq. (\ref{Eq:35}) in Figure {\ref{fig:14}}.

In particular, when it satisfies the relation of both $4U^{2}/\kappa_{zz}=0$ and $t_{A}=t_{D}/2$, the interplay of the cooling, escape, and systematic energy gain from shock acceleration in the downstream radiation zone tends to establish a power-law distribution of emission electrons in the energy range of $\gamma_{0}\leq\gamma\leq\gamma_{max}$ for a continuous mono-energetic injection. The radiative cooling and acceleration restrict the particle energy neither above $\gamma_{max}$ nor under $\gamma_{0}$.

\subsection{Spatially dependent transport equation}					
Here, we solve Eqs. (\ref{Eq:7}), (\ref{Eq:13}), and (\ref{Eq:15}), which include non-relativistic shock acceleration and relativistic shock acceleration, and discuss the solution of Eq. (\ref{Eq:7}) in the presence of radiative cooling. We consider a plane-geometric shock propagating along with the \textit{z} axis with speed $U$. The shock front is perpendicular to both the \textit{z} axis at $z=0$ and the magnetic field $B_{0}$ in the shock frame. All the velocities below subscript 1 refer to upstream and those below subscript 2 to downstream.

{\bf \emph{Case (9): Non-relativistic diffusive shock acceleration}}~~In this case, we assume $Q = 0$ in Eq. (\ref{Eq:7}) and treat the downstream region as infinite. In the boundary conditions $F(z>0, p)=F_{1}(p)\propto\delta(p-p_{0})$ and $S_{flux}=\kappa_{1}\frac{\partial}{\partial z}F_1(z>0, p)=\kappa_{2}\frac{\partial}{\partial z}F_2(-\infty,p)=0$,  the downstream stationary solution shows a power-law distribution (Drury 1983; Dermer\textit{ et al}. 2009; Jones \textit{et al}. 1991),
\begin{eqnarray}
F_{2}(-\infty<z<0, p)&=&Mp^{-M}\int_{p_{1}}^{p}F_{1}(p^{'})~p^{'(M-1)}~dp^{'}\nonumber\\&+&c_{1}p^{-M} \propto p^{-M}\;,
\label{Eq:36}
\end{eqnarray}
where the index $M=3r/(r-1)$ with the compression ratio $r\simeq U_{2}/U_{1}$, $c_{1}$ is an arbitrary integral constant, and $p_{1}$ satisfies $p_{1}/m_{e}\gamma=v_{1}>>U$. These results hold with the assumption of the diffusion approximation and can be found in the literature on non-relativistic diffusive shock acceleration (e.g., Krymsky \textit{et al}. 1977; Axford \textit{et al}. 1977; Bell 1978; Drury 1983; Blandford \textit{et al.} 1987; Jones \textit{et al}. 1991). It can be seen that the spectral shape depends on the compression ratio $r$ rather than on both the scattering process and the shock geometry. This is an advantage of test particles for shock acceleration in astrophysics.
	
In particular, we can prove $\frac{3}{r-1}\approx\frac{t_{sh}}{T_{esc}}$ with an individual particle kinetic approach (Protheroe \textit{et al}. 2004), where ${T_{esc}}$ is the escape timescale. Since
\begin{eqnarray}
n(p)&=&4\pi p^{2}F_{2}(p)\propto p^{-(r+2)/(r-1)}\nonumber\\&=&p^{-3/(r-1)-1}\approx p^{-(t_{sh}/T_{esc}+ 1)}\;,
\label{Eq:37}
\end{eqnarray}
if we assume $T_{esc}=t_{esc}$ and $D(p)=0$, then $M=-(t_{sh}/T_{esc}+1)$ equals the spectral index in Eq. (\ref{Eq:33}). It can be seen that these results come from the shock acceleration and particle escapes. Note that these approximative results are obtained in the framework of test particles, and the non-linear effects of energetic particle pressure on the shock profile can induce the spectrum to harden (Demer \textit{et al}. 2009). This is beyond the scope of the present work.
 									
\begin{figure}
  \centering
  \includegraphics[width=9 cm]{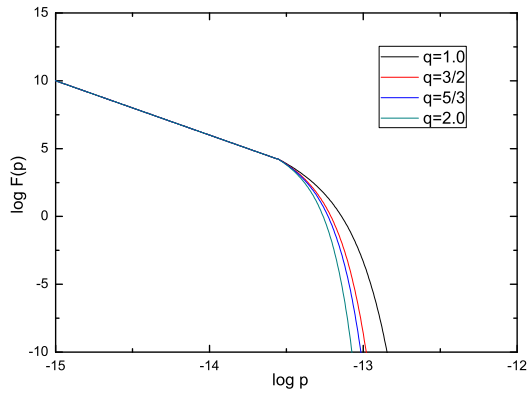}\\
  \caption{EED as given in Eq. (\ref{Eq:38}) at $q=1.0$, $q=3/2$, $q=5/3$, and $q=2.0$ with compression ratio $r=4$.}\label{fig:15}
\end{figure}

{\bf \emph{Case (10): Non-relativistic diffusive shock acceleration, synchrotron or/ and IC cooling, and mono-energetic injection}}~~In this case, we assume that mono-energetic electron populations with a spectrum $Q\propto\delta(p-p_{0})\delta(z-z_{0})/4\pi p^{2}$ are injected. Adding a radiative term $\frac{\partial}{\partial p}( \eta p^{4}F)/p^{2}$ to Eq. (\ref{Eq:7}), this transport equation can describe the shock acceleration and radiation. The analytical steady-state solution can be obtained either in a limited region of momentum space at the shock front (Webb \textit{et al}. 1984; Zirakashvili \textit{et al}. 2007; Vannoni \textit{et al}. 2009) or for some specific choices of the diffusion coefficient (Blasi 2010). The analytical or semi-analytical investigation (Webb \textit{et al}. 1984; Zirakashvili \textit{et al}. 2007; Blasi 2010) for an arbitrary diffusion coefficient indicate that 1) in the lower-momentum regime, the spectrum keeps the standard test-particle distribution as Eq. (\ref{Eq:36}); 2) when $r>4$ is satisifed, the spectrum shows a small hump near the equilibrium momentum $p_{eq}$ that is determined by $t_{loss}(p)=t_{A}(p)$ when the stochastic acceleration is neglected; and 3) in the momentum $p>p_{eq}$ regime, if the energy-dependent diffusion coefficient can be written as a general form $\kappa_{zz}(p)\propto p^{q}$, the radiative cooling should influence the spectral shape with an exponential term $\exp[-(p/p_{eq})^{(q+1)}]$. These results can be expressed as
\begin{eqnarray}
F(p) \propto \left\{
\begin{array}{ll}
p^{-M}\;,~~~p\leq p_{eq}\\
p^{-M}\exp \left[-(p/p_{eq})^{(q + 1)}\right]\;,~~~p>p_{eq}
\end{array} \right.\;.
\label{Eq:38}
\end{eqnarray}
We show the EED as given in Eq. {\ref{Eq:38}} at $q=1.0$, $q=3/2$, $q=5/3$, and $q=2.0$ in Figure {\ref{fig:15}}.
Note that the ratio	of the magnetic strength between in the downstream and in the upstream region can also influence the spectral shape.

{\bf \emph{Case (11): Relativistic shock acceleration}}~~Owing to the high anisotropy, an analytic solution for Eq. (\ref{Eq:15}) has not yet been available, while the analytic expression for the spectral index could be obtained (Sironi 2015) by the analytical methods, such as approximately matching numerical eigenfunctions of the transport equation between the upstream region and the downstream region (Kirk and Schneider 1987; Kirk \textit{et al}. 2000), the distribution expansion parallel to the shock front (Keshet and Waxman 2005), and Monte Carlo simulations (Bednarz and Ostrowski 1998; Achterberg \textit{et al}. 2001; Ellison \textit{et al}. 2013). Expanding the distribution about the variable $\mu+U/c$ for isotropic diffusion with the boundary conditions $f_{1}(\mu_{1}, p_{1}, \tau=0)=f_{2}(\mu_{2}, p_{2}, \tau=0)$ and $f_{1}(\mu_{1}, p_{1}, \tau\to  -\infty)=0$, $f_{2}(\mu_{2}, p_{2}, \tau\to\infty)\propto p_{2}^{-s_{2}}$, we can find (Keshet and Waxman 2005)
\begin{equation}
f_2\left( p_2 \right) \propto {p_2^{ - {s_{iso}}}}\;,
\label{Eq:39}
\end{equation}
where $s_{iso}=(3\beta_{1}-\beta_{1}\beta_{2}^{2}+\beta_{2}^{3})/(\beta_{1}-\beta_{2})$ is a simple expression for $s_2$, $\beta_{1,2}=U_{1,2}/c$. For ultra-relativistic shock, $s_{iso}(\beta_{1}\to1, \beta_{2}\to1/3)=38/9\approx 4.222$ is in agreement with the numerical simulation result, where it is suggested that $4.2\le s_{iso}\le4.4$.

\section{Summary and Discussion}
It is considered that the EED encodes important information about particle acceleration and cooling, jet physics, and the environment. To determine the EED shape in a jet emission region, there is great interest in the inversion of the jet physics and environment from the EED formation. In this paper, we describe the theory of the \emph{Fermi} acceleration and the distribution of electron spectrum. Including the theory of the turbulent acceleration and the derivation of general F-P equation, we take into account the basic particle transport equation and the F-P equation with spatially homogeneous isotropic distribution. In the frame of several special physical processes, we construct the particle transport equation, and obtain the analytical or semi-analytical solution. We summarize the electronic spectrum and the formation mechanism as follows:\\
$\bullet$ In the case of turbulence acceleration with \textit{q}=2, escape, and mono-energetic injection, we expect a double power-law spectrum, Eq. (\ref{Eq:19}), in the steady-state, and a time-dependent log-parabolic spectrum, Eq. (\ref{Eq:20}) (case 1);\\
$\bullet$ In the case of turbulence acceleration with \textit{q}=2, and initial mono-energetic injection, we expect a time-dependent log-parabolic spectrum, Eqs. (\ref{Eq:21}) and   (\ref{Eq:22}) (case 2);\\
$\bullet$ In the case of turbulence acceleration with \textit{q}=2, initial mono-energetic injection, and radiative cooling, we expect a time-dependent log-parabolic spectrum, Eq. (\ref{Eq:23}), and a Maxwellian distribution in the steady state, Eq. (\ref{Eq:24}) (case 3);\\
$\bullet$ In the case of turbulence acceleration with \textit{q}=2, escape, continuous mono-energetic injection, and radiative cooling, we expect a power-law spectrum in lower-energy regimes and a power-law spectrum with a exponential cutoff in higher-energy regimes, Eq. (\ref{Eq:25}). If in the case of continuous injection at low energy, a double power-law spectrum in lower-energy regimes is shown. If in the case of inefficient escape,  a modified Maxwellian distribution in higher-energy regimes, Eq. (\ref{Eq:26}), is shown (case 4);\\
$\bullet$ In the case of radiative cooling and a constant escape timescale in the downstream region with a steady power-law injection near the shock surface, we expect a double power-law spectrum, Eq. (\ref{Eq:29}) (case 5);\\
$\bullet$ In the case of radiative cooling in the downstream region with a steady power-law injection near the shock surface,  we expect a double power-law spectrum, Eqs. (\ref{Eq:31}) and (\ref{Eq:32}) (case 6);\\
$\bullet$ In the case of a constant escape, second-order \emph{Fermi} acceleration, we expect a power-law spectrum, Eq. (\ref{Eq:33}), in the steady state (case 7);\\
$\bullet$ In the case of radiative cooling, escape, and first-order \emph{Fermi} acceleration with a constant mono-energetic injection, we expect a power-law spectrum, Eq. (\ref{Eq:35}) (case 8);\\
$\bullet$ In the case of only non-relativistic diffusive shock acceleration, we expect a power-law spectrum, Eq. (\ref{Eq:36}) (case 9);\\
$\bullet$ In the case of non-relativistic shock acceleration with synchrotron radiative cooling, we expect a power-law spectrum in lower-energy regimes with a spectral index of approximately 2, and a power-law spectrum with an exponential cutoff in higher-energy regimes, Eq. (\ref{Eq:38}) (case 10);\\
$\bullet$ In the case of only relativistic shock acceleration, we expect a power-law spectrum, Eq. (\ref{Eq:39}) (case 11).

Traditionally, we can employ one of the EEDs in the above 11 cases to reproduce the multi-wavelength emission of a blazar. If the SED could be reproduced well, we can invert the jet physics and environment from the EED shape. In these approaches, the essential difference between spectral intensities and spectral fluences are not taken into account.

The recent investigation suggested that there is a completely different spectral shape of the EED at a given time and a time-integrated EED (Zacharias and Schlickeiser 2009; Zacharias and Schlickeiser 2010). After further discussion, it was concluded if the flaring timescale is much shorter than the integration timescale, one can use total electron fluences (2012a; 2012b). Actually, high-energy observations of a blazar use very long integration times of the order of days and weeks to collect statistically enough photons from a flaring AGN source. These integration times are orders of magnitude longer than both the radiative loss times and the acceleration timescales of the radiating particles. In such cases we cannot use the EED at a given time to compare with the observations; instead, we must use a time-integrated EED.

However, a major complication arises as observations at different frequencies using different integration times when observing a blazar. At radio frequencies, short (minutes) snapshots are used, whereas at very high frequencies, very long integration times are used. These are then plotted in one SED, which makes the analysis very difficult.


\acknowledgments
\section*{Acknowledgements}
We thank the anonymous referee for valuable comments and suggestions.
This work was partially supported by the National Natural Science Foundation of China (Grant Nos. 11133006, 11433004, 11463007 and 11673060), and the Natural Science Foundation of Yunnan Province (Grant Nos. 2016FB003 and 2017FD072). Additional support was provided by the Specialized Research Fund for Shandong Provincial Key Laboratory and by the Key Laboratory of Particle Astrophysics of Yunnan Province (Grant No. 2015DG035).

\clearpage


\begin{thebibliography}{}

\bibitem[Abdo et al. (2010)]{ABDO10}
Abdo, A.A., Ackermann, M., Agudo, I. et al., 2010, \apj, 716, 30
\bibitem[Abdo et al. (2011a)]{ABDO11a}
Abdo, A.A., Ackermann, M., Ajello, M. et al., 2011a, \apj, 727, 129
\bibitem[Abdo et al. (2011b)]{ABDO11b}
Abdo, A.A., Ackermann, M., Ajello, M. et al., 2011b, \apj, 736, 131
\bibitem[Achatz et al. (1991)]{AC91}
Achatz, U., Steinacker, J., \& Schlickeiser, R., 1991, \aap, 250, 266
\bibitem[Ackermann et al. (2011)]{AC11}
Ackermann, M., Ajello, M., Allafort, A. et al., 2011, \apj, 743, 171
\bibitem[Ackermann et al. (2015)]{AC15}
Ackermann, M., Ajello, M., Atwood, W., et al., 2015, \apj, 810, 14
\bibitem[Achterberg et al. (2001)]{AC01}
Achterberg, A., Gallant, Y. A., Kirk, J. G. \& Guthmann, A.W.: 2001, \mnras, 328, 393
\bibitem[Aharonian et al. (2009)]{AH09}
Aharonian, F., Akhperjanian,A. G., Anton,G.,et al., 2009, \apj, 696, L150
\bibitem[Albert et al. (2007)]{AL07}
Albert, J. et al., 2007, \apj, 669, 862 (2007)
\bibitem[Asano et al. (2014)]{AS14}
Asano, K., Takahara, F., Kusunose, M., Toma, K., \& Kakuwa, J.: 2014, \apj, 780, 64
\bibitem[Axford et al. (1977)]{AX77}
Axford, W.I., Lear, E., \& Skadron, G., 1977, in Proc. 15th Int. Cosmic Ray Conf., Plovdiv vol.11, p 132
\bibitem[Becker et al. (2006)]{BE06}
Becker, P.A., Le, T., \& Dermer, C.D., 2006, \apj, 647, 539 (2006)
\bibitem[Bednarz et al. (1998)]{ABDO09a}
Bednarz, J., \& Ostrowski, M., 1998, \prl, 80, 3911
\bibitem[Bell (1978)]{BE78}
Bell, A. R., 1978, \mnras, 182, 147
\bibitem[Blandford et al. (1978)]{BL78}
Blandford, R.D., \& Ostriker, J.P., 1978, \apj, 221, 29
\bibitem[Blandford et al. (1987)]{BL87}
Blandford, R., \& Eichler, D., 1987, \physrep, 154, 1
\bibitem[Blasi (2010)]{BL10}
Blasi, P., 2010, \mnras, 402, 2807
\bibitem[Bottcher et al. (2002)]{BO02}
B$\rm\ddot{o}$ttcher, M., \& Dermer, C. D.: 2002, \apj, 564, 86
\bibitem[Chen (2014)]{CH14}
Chen, L., 2014, \apj, 788, 179
\bibitem[Chiaberge et al. (1999)]{CH99}
Chiaberge, M., \& Ghisellini, G., 1999, \mnras, 306, 551
\bibitem[Dermer (2009)]{DE09}
Dermer, C.D., \& Menon, G., 2009, High Energy Radiation from Black Holes., Prinseton university press, Prinseton and oxfod
\bibitem[Drury et al. (1983)]{DR83}
Drury, L.: 1983, \ssr, 36, 57
\bibitem[Dung \& Schlickeiser (1990a)]{DS90a}
Dung, R., \& Schlickeiser, R., 1990a, \aap, 237, 504
\bibitem[Dung \& Schlickeiser (1990b)]{DS90b}
Dung, R., \& Schlickeiser, R., 1990b, \aap, 240, 537
\bibitem[Ellison et al. (2013)]{EL13}
Ellison, D.C., Warren, D.C., Bykov, A.M., 2013, \apj, 776, 46
\bibitem[Fermi (1949)]{FE49}
Fermi, E., 1949, Phys. Rev., 75, 1169
\bibitem[Fermi (1954)]{FE54}
Fermi, E., 1954, \apj, 119, 1
\bibitem[Finke et al. (2008)]{FI08}
Finke, J.D., Dermer, C.D., \& B$\ddot{o}$ttcher, M., 2008, \apj, 686, 181
\bibitem[Fink (2013)]{FI13}
Finke, J.D., 2013, \apj, 763, 134
\bibitem[Finke et al. (2014)]{FI14}
Finke, J.D., \& Becker, P. A., 2014, \apj, 791, 21
\bibitem[Fossati et al. (1998)]{FO98}
Fossati, G., Maraschi, L., Celotti, A. et al., 1998, \mnras, 299, 433
\bibitem[Hayashida et al. (2012)]{HA12}
Hayashida, M., Madejski, G. M., Nalewajko, K. et al., 2012, \apj, 754, 114
\bibitem[Jokipii (1966)]{JO66}
Jokipii, J.R., 1966, \apj, 143, 961
\bibitem[Jokipii (2001)]{Jo01}
Jokipii, J.R., 2001, \apss, 277, 15
\bibitem[Jones et al. (1991)]{JO91}
Jones, F.C., \& Ellison, D.C., 1991, \ssr, 58, 259
\bibitem[Jones (1994)]{JO94}
Jones, F.C., 1994, \apjs, 90, 561
\bibitem[Kakuwa et al. (2015)]{KA15}
Kakuwa, J., Toma, K., Asano, K., Kusunose, M., \& Takahara, F., 2015, \mnras, 449, 551
\bibitem[Kardashev (1962)]{KAR62}
Kardashev, N.S., 1962, \sovast, 6, 317
\bibitem[Kataoka et al. (2000)]{KA00}
Kataoka, J., Takahashi, T., Makino, F. et al., 2000, \apj, 528, 243 (2000)
\bibitem[Katarzynski et al. (2001)]{KA01}
Katarzynski, K., Sol, H., \& Kus, A., 2001, \aap, 367, 809
\bibitem[Katarzynski et al. (2006)]{KA06}
Katarzynski, K., Ghisellini, G., Mastichiadis, A., Tavecchio, F., \& Maraschi, L., 2006, \aap, 453, 47
\bibitem[Keshet et al. (2005)]{KE05}
Keshet, U., \& Waxman, E., 2005, \prl, 94, 111102
\bibitem[Kirk et al. (1987)]{KI87}
Kirk, J.G., \& Schneider, P., 1987, \apj, 315, 425
\bibitem[Kirk et al. (1988)]{KI88}
Kirk, J.G., Schneider, P., \& Schlickeiser, R., 1988, \apj, 328, 269
\bibitem[kirk et al. (1992)]{KI92}
Kirk, J.G. \& Wassmann, M., 1992, \aap, 254, 167
\bibitem[Kirk et al. (1998)]{KI98}
Kirk, J.G., Rieger, F.M., \& Mastichiadis, A., 1998, \aap, 333, 452
\bibitem[Kirk et al. (2000)]{KI00}
Kirk, J.G., Guthmann, A.W., Gallant, Y.A., \& Achterberg, A., 2000, \apj, 542, 235
\bibitem[Krymsky et al. (1977)]{KR77}
Krymsky, G.F., 1977, Dokl. Akad. Nauk. SSSR, 243, 1306
\bibitem[Lee (1982)]{LEE82}
Lee, M.A., 1982, \jgr, 87, 5063
\bibitem[Longair (2015)]{LIU15}
Liu, S.M., 2015, Sci Sin-Phys Mech Astron, 45, 119509
\bibitem[Malkov et al. (2001)]{MA01}
Malkov, M.A., \& Drury, L. O'C., 2001, Reports on Progress in Physics, 64, 429
\bibitem[Massaro et al. (2004)]{MA04}
Massaro, E., Perri, M., Giommi, P., \& Nesci, R., 2004, \aap, 413, 489
\bibitem[Massaro et al. (2006)]{MA06}
Massaro, E., Tramacere, A., Perri, M., Giommi, P., \& Tosti, G., 2006, \aap, 448, 861
\bibitem[Mertsch (2011)]{ME11}
Mertsch, P., 2011, JCAP, 12, 10
\bibitem[Miller et al. (1995)]{MI95}
Miller, J.A., \& Roberts, D.A., 1995, \apj, 452, 912
\bibitem[Paggi et al. (2010)]{PA10}
Paggi, A., 2010, Frares in Blazars. Rome:degli di Rome Universita
\bibitem[Park et al. (1995)]{PA95}
Park, B.T., \& Petrosian, V., 1995, \apj, 446, 699
\bibitem[Parker (1965)]{PAR65}
Parker, E.N., 1965, \ssr, 4, 666
\bibitem[Petrosian (2004)]{PE04}
Petrosian, V., \& Liu S., 2004, \apj, 610, 550
\bibitem[Petrosian (2012)]{PE12}
Petrosian, V., 2012, \ssr, 173, 535
\bibitem[Protheroe et al. (2004)]{PR04}
Protheroe, R.J., \& Clay, R.W., 2004, \pasp, 21, 1
\bibitem[Rieger et al. (2007)]{RI07}											
Rieger, F. M., Bosch-Ramon, V., \& Duffy, P., 2007, \apss, 309, 119
\bibitem[Roelof (1969)]{RO69}
Roelof, E.C., 1969, Lectures in High Energy Astrophysics, eds. H. $\rm \ddot{O}$gelman and J.R. Wayland, Vol. 111 (Washington: NASA SP-199)
\bibitem[Schlickeiser (1984)]{SC84}
Schlickeiser, R., 1984, \aap, 136, 227
\bibitem[Schlickeiser (1985)]{SC85}
Schlickeiser, R., 1985, \aap, 143, 431
\bibitem[Schlickeiser et al.(1987)]{SC87}
Schlickeiser, R., Sievers, A., \& Thiemann, H., 1987, \aap, 182, 21
\bibitem[Schlickeiser (1989)]{SC89}
Schlickeiser, R., 1989, \apj, 336, 243
\bibitem[Schlickeiser et al. (2000)]{SC00}
Schlickeiser, R., \& Dermer, C.D., 2000, \aap, 360, 789
\bibitem[Schlickeiser (2002)]{SC02}
Schlickeiser, R., 2002, Cosmic Ray Astrophysics (Berlin: Springer)
\bibitem[Schlickeiser (2015)]{SC15}
Schlickeiser, R., 2015, \apj, 809, 124
\bibitem[Schneider et al. (1989)]{SC89}
Schneider, P., \& Kirk. J.G., 1989, \aap, 217, 344
\bibitem[Sironi (2015)]{SI15}
Sironi, L., \& Lemoine, M., 2015, \ssr, 191, 519
\bibitem[Skilling (1975)]{SK75}
Skilling, J., 1975, \mnras, 172, 557
\bibitem[Somov (2013)]{SO13}
Somov, B.V.,  2013, Plasma Astrophysics, Part I. Springer New York Heidelberg Dordrecht.London
\bibitem[Spitkovsky (2008a)]{SP08a}
Spitkovsky, A., 2008, \apj, 673, L39
\bibitem[Steinacker et al. (1992)]{ST92}
Steinacker, J., \& Miller, J.A., 1992, \apj, 393, 764
\bibitem[Stawarz et al. (2008)]{ST08}
Stawarz, L., \& Petrosian,V., 2008, \apj, 681, 1725
\bibitem[Tademaru (1969)]{TA69}
Tademaru, E., 1969, \apj, 336, 264
\bibitem[Tavecchio et al. (1998)]{ABDO09a}
Tavecchio, F., Maraschi L., \& Ghisellini, G., 1998, \apj, 509, 608
\bibitem[Tavecchio et al. (2010)]{TA10}
Tavecchio, F., Ghisellini G., Ghirlanda G., Foschini L., Maraschi L., 2010, \mnras, 401, 1570
\bibitem[Tramacere et al. (2009)]{TR09}
Tramacere, A., Giommi, P., Perri, M., Verrecchia, F., \& Tosti, G., 2009, \aap, 501, 879
\bibitem[Tramacere et al. (2011)]{TR11}
Tramacere, A., Massaro, E., \& Taylor, A. M., 2011, \apj, 739, 66
\bibitem[Urry et al. (1995)]{UR95}
Urry, C.M., \& Padovani, P., 1995, \pasp, 107, 803
\bibitem[Ushio et al. (2010)]{US10}
Ushio, M., Stawarz, L., Takahashi, T. et al., 2010, \apj, 724, 1509
\bibitem[Vannoni et al. (2009)]{VA09}
Vannoni, G., Gabici, S., \& Aharonian, F. A., 2009, \aap, 497, 17
\bibitem[Verkhoglyadova et al. (2014)]{VE14}
Verkhoglyadova, O.P., Zank, G.P., Li, G., 2015, \physrep, 557, 1
\bibitem[Virtanen et al. (2005)]{VI05}
Virtanen, J.J.P., \& Vainio, R., 2005, \apj, 621, 313
\bibitem[Webb et al. (1984)]{WE84}
Webb, G.M., Drury, L.O'C., Biermann, P., 1984, \aap, 137, 185
\bibitem[Webb (1985)]{WE85}
Webb, G.M., 1985, \apj, 296, 319
\bibitem[Yan et al. (2013)]{YA13}
Yan, D., Zhang, L., Yuan, Q., Fan, Z., \& Zeng, H., 2013, \apj, 765, 122
\bibitem[Zacharias \& Schlickeiser (2009)]{ZS09}
Zacharias, M., \& Schlickeiser, R., 2009, \aap, 498, 667
\bibitem[Zacharias \& Schlickeiser (2010)]{ZS10}
Zacharias, M., \& Schlickeiser, R., 2010, \aap, 524, A31
\bibitem[Zacharias \& Schlickeiser (2012a)]{ZS12a}
Zacharias, M., \& Schlickeiser, R., 2012a, \apj, 761, 110
\bibitem[Zacharias \& Schlickeiser (2012b)]{ZS12b}
Zacharias, M., \& Schlickeiser, R., 2012b, \mnras, 420, 84
\bibitem[Zheng \& Zhang (2011a)]{ZH11}
Zheng, Y.G., \& Zhang, L., 2011a, \apj, 728, 105
\bibitem[Zheng \& Zhang (2011b)]{ZH11}
Zheng, Y.G., \& Zhang, L., 2011b, 32ND International Cosmic Ray Conference, Beijing 2011, 8, 175
\bibitem[Zheng et al. (2018)]{ZH18}
Zheng, Y.G., Long, G.B., Yang, C.Y., \& Bai, J.M., 2018, accepted in \mnras
\bibitem[Zirakashvili et al. (2007)]{ZI07}
Zirakashvili, V.N., \& Aharonian, F., 2007, \aap, 465, 695

\end{thebibliography}
\end{document}